\documentclass[10pt,journal,compsoc]{IEEEtran}

\ifCLASSOPTIONcompsoc
  \usepackage[nocompress]{cite}
\else
  \usepackage{cite}
\fi

\usepackage
{hyperref}

\usepackage{graphicx}
\graphicspath{{./figs/}}

\usepackage{mathtools}
\usepackage{amssymb}
\usepackage{bm}
\usepackage{wrapfig}
\usepackage{hhline}
\usepackage[usenames,dvipsnames,svgnames,table]{xcolor}
\usepackage{soul}
\usepackage{comment}

\usepackage[ruled,linesnumbered]{algorithm2e}

\usepackage{dsfont}
\usepackage{subfig}

\usepackage[mathletters]{ucs}
\usepackage[utf8x,utf8]{inputenc}
\usepackage{upgreek}
\usepackage{dblfloatfix}

\usepackage{microtype}

\usepackage{booktabs}

\pdfpageattr {/Group << /S /Transparency /I true /CS /DeviceRGB>>}

\newif\ifsubmit
\submittrue
\ifsubmit
    \newcommand{\todo}[1]{}
    \newcommand{\TODO}[1]{}
    \newcommand{\MAYBEDO}[1]{}
    
    \newcommand{\yotam}[1]{}
    \newcommand{\steve}[1]{}
    \newcommand{\jingwan}[1]{}
    \newcommand{\jianchao}[1]{}
    
    \newcommand{\topic}[1]{}
    
    \newcommand{\rev}[1]{#1}
    \newcommand{\revtwo}[1]{#1}
    \newcommand{\revthree}[1]{#1}
\else
    \newcommand{\todo}[1]{\textcolor{red}{TODO: #1}}
    
    \newcommand{\TODO}[1]{\textcolor{red}{\textbf{****** #1 ******}}}
    \newcommand{\MAYBEDO}[1]{\textcolor{brown}{\textbf{****** \textsc{(maybe)} #1 ******}}}
    
    \newcommand{\yotam}[1]{{\color{purple}\textsc{Yotam:} #1}}
    \newcommand{\steve}[1]{{\color{green}\textsc{Steve:} #1}}
    
    \newcommand{\jianchao}[1]{{\color{orange}\textsc{Jianchao:} #1}}
    
    \newcommand{\topic}[1]{{\color{magenta}\textsc{Topic:} #1\\}}
    \newcommand{\rev}[1]{#1}
    \newcommand{\revtwo}[1]{#1}
    \newcommand{\revthree}[1]{\textcolor{blue}{#1}}

\fi

\usepackage{letltxmacro}
\LetLtxMacro{\oldmarginpar}{\marginpar}
\renewcommand{\marginpar}[1]{\oldmarginpar{\color{red}\tiny #1}}
\DeclareMathOperator{\softmax}{softmax}
\DeclareMathOperator{\argmin}{argmin}
\newcommand*{\x}{\mathsf{x}\mskip1mu}

\newcommand{\shortcite}[1]{\cite{#1}}
\renewcommand{\paragraph}[1]{\textbf{#1} }

\begin{document}


\title{Pigmento: Pigment-Based Image Analysis and Editing}


\author{Jianchao~Tan, Stephen~DiVerdi, Jingwan~Lu, Yotam~Gingold
\IEEEcompsocitemizethanks{\IEEEcompsocthanksitem J.~Tan and Y.~Gingold are with George Mason University.\protect\\
\IEEEcompsocthanksitem S.~DiVerdi and J.~Lu are with Adobe Research.}
}

\markboth{Pigmento: Pigment-Based Image Analysis and Editing}%
{Pigmento: Pigment-Based Image Analysis and Editing}


\IEEEtitleabstractindextext{%
\begin{abstract}
%
The colorful appearance of a physical painting is determined by the distribution of paint pigments across the canvas, which we model as a per-pixel mixture of a small number of pigments with multispectral absorption and scattering coefficients.  We present an algorithm to efficiently recover this structure from an RGB image, yielding a plausible set of pigments and a low RGB reconstruction error.  We show that under certain circumstances we are able to recover pigments that are close to ground truth, while in all cases our results are always plausible. Using our decomposition, we repose standard digital image editing operations as operations in pigment space rather than RGB, with interestingly novel results.  We demonstrate tonal adjustments, selection masking, cut-copy-paste, recoloring, palette summarization, and edge enhancement.

\end{abstract}

}

\maketitle

\IEEEdisplaynontitleabstractindextext

%
\IEEEpeerreviewmaketitle

%


\begin{figure*}
\centering
\includegraphics[width=0.95\textwidth]{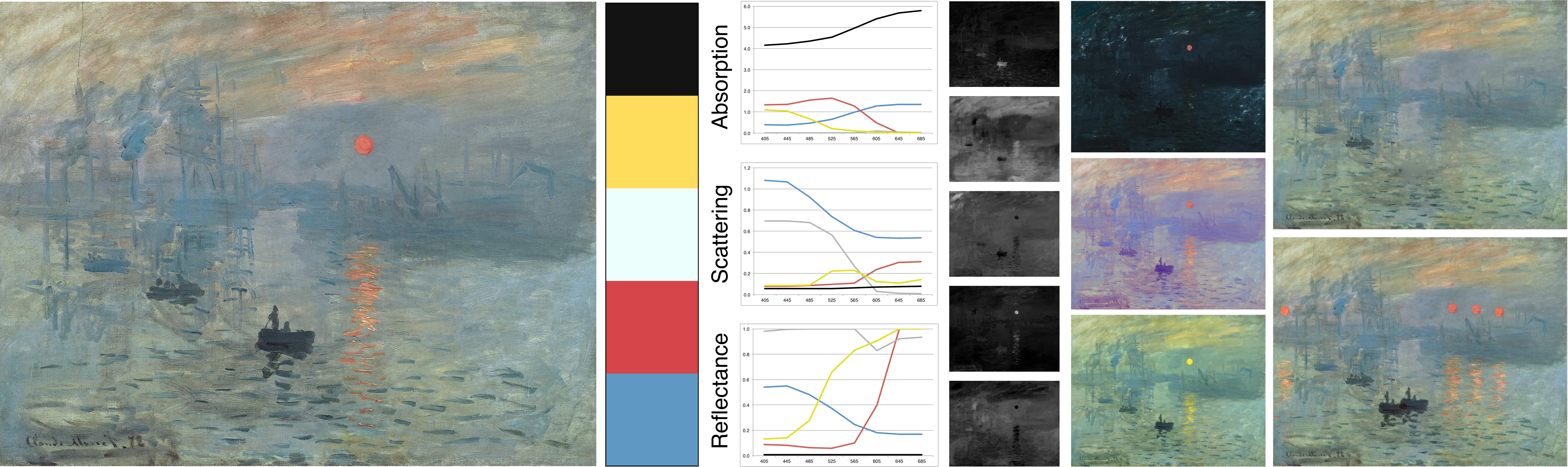}
\caption{Analysis and editing of Monet's ``Impression, soleil levant.''  From left to right, input image, extracted palette in RGB, multispectral coefficient curves for palette pigments, mixing weights, recoloring, and cut-copy-paste.}
\label{fig:teaser}
\end{figure*}


\section{Introduction}
\label{sec:intro}

%
Stated generally, a ``painting'' in the physical world is a two-dimensional arrangement of material.
This material may be oil or watercolor paint, or it may be ink from a pen or marker, or charcoal or pastel.
These pigments achieve a colorful appearance by virtue of how they absorb and reflect light and their thickness.
%
Kubelka and Munk \cite{kubelka1931article,kubelka1948new} described a model for the layering of physical materials,
and Duncan \cite{Duncan:1940:CPM} extended it to include homogeneous mixing.
In this model, the appearance of a material (reflectance and transmission of light) is defined by how much it scatters and absorbs each wavelength of light and its overall thickness.
These models are widely used to model the appearance of paint, plastic, paper, and textiles;
they have been used previously in the computer graphics literature
\cite{Curtis:1997:CW,Baxter:2004:IMPaSTo,Lu:2014:RPC,Tan:2015:DTL}.

When painting, artists choose or create
a relatively small set of pigments to be used throughout the painting.
We call this set the \emph{primary pigment palette}.
We assume that all observed colors in the painting are created by mixing or layering
pigments from the palette.

When we view a painting, either directly with our eyes or indirectly after digitizing it into a three-channel RGB image,
we observe only the overall reflectance and not the underlying material parameters.
In RGB-space, the underlying pigments which combine to form the appearance of a pixel are not accessible for editing.
One color in the palette cannot be easily changed or replaced.
Translucent objects, common in paintings due to the mixing of wet paint,
cannot be easily extracted or inserted.

We propose an approach to decompose a painting into its constituent pigments in two stages. First, we compute a small set of pigments in terms of their Kubelka-Munk (KM) scattering and absorption parameters.
Second, we compute per-pixel mixing proportions for the pigments that reconstruct the original painting.  We show that this decomposition has many desirable properties.  Particularly for images of paintings, it is able to achieve lower error reconstructions with smaller palettes than previous work.  Furthermore, the decomposition enables image editing applications to be posed in pigment space rather than RGB space, which can make them more effective or more expressive.  We demonstrate tonal adjustments by editing pigment properties; recoloring; selection masking; copy-paste; palette summarization; and edge enhancement.


Thematically, this work is similar to Lillicon~\cite{Bernstein:2015:LUT} and Project Naptha~\cite{naptha}, which both present ways to interpret structure in unstructured documents to enable high level edits based on the interpreted structure.  In Lillicon's case, the structure is an alternate vector representation of the artwork, while in Project Naptha, the structure is styled text within the image.  Our contribution is to apply this strategy to flat, unstructured RGB images of paintings, which are created via a complex structure (physical pigments and brush strokes).  Our analysis allows us to interpret the complex structure of the painting from the RGB image, which enables editing operations based on that structure.

\section{Related Work}
\label{sec:related}

Our work is inspired by the recent efforts of Tan et al.~\shortcite{Tan:2016:DIL}, Lin et al.~\shortcite{lin2017layer}, Aksoy et al.~\shortcite{aksoy2017unmixing} and Zhang et al.~\shortcite{zhang2017palette} to decompose an arbitrary image into a small set of partially transparent layers suitable for RGB compositing. 
Tan et al.~\shortcite{Tan:2016:DIL} use RGB-space convex hull geometry to extract a palette, and then solve an optimization problem to extract translucent layers for the Porter-Duff ``over" compositing operator (alpha compositing), which is the standard color compositing model.
Lin et al.~\shortcite{lin2017layer} extract translucent layers from images and videos based on an additive color mixing model. They use locally linear embedding, which assumes that each pixel is a linear combination of its neighbors. 
Aksoy et al.~\shortcite{aksoy2017unmixing} extract translucent layers from images, also based on an additive color mixing model. However, unlike Tan et al.~\shortcite{Tan:2016:DIL} and Lin et al.~\shortcite{lin2017layer},
each layer's color varies spatially.
Zhang et al.~\shortcite{zhang2017palette} use a clustering-based method to extract palette colors and then decompose the entire image into a linear combination of them.
This is a similar representation as the additive mixing layers from Lin et al.~\shortcite{lin2017layer} and Aksoy et al.~\shortcite{aksoy2017unmixing}. 
All of these decompositions allow users to edit the image in a more intuitive manner, effectively segmenting the image by color and spatial coherence.
Similarly, Chang et al.~\shortcite{Chang:2015:PPR} extract a small palette of colors from an image and implicitly model each pixel as a mixture of those palette colors to enable image recoloring using radial basis functions.
We extend these results specifically for physical paintings by using a physically-inspired model of pigment mixing (Kubelka-Munk) and estimating multispectral (greater than RGB) pigment properties.

Our work is contemporaneous with A\rev{har}oni-Mack et al.~\shortcite{aharoni2017pigment},
who decompose watercolor paintings into linear mixtures of a small set of primary pigments \rev{also using the Kubelka-Munk mixture model.
The primary differences between our approaches is that they target (translucent) watercolor paintings
and use 3-wavelength (RGB) parameters with varying thickness,
while we evaluate our approach with (opaque) acrylic and oil paintings and compute an 8-wavelength constant-thickness decomposition.
They similarly use a convex-hull in color-space to identify palette colors.
Both methods regularize the problem at least in part with spatial smoothness.
Both methods leverage existing datasets of measured Kubelka-Munk scattering and absorption parameters (3-wavelength watercolor pigment parameters from Curtis et al.~\shortcite{Curtis:1997:CW} versus 33-wavelength acrylic parameters from Okumura~\cite{okumura2005developing}).}

Algorithmically, our work is most similar to that of Kauvar et al.~\shortcite{Kauvar:2015:ACD}, which optimizes a set of multispectral illuminants and linear mixing weights to reproduce an image. This is suitable for their scenario (choosing projector illuminants) but not for mimicking physical paintings. The nonlinear nature of the Kubelka-Munk equations makes our problem much harder.

While the Kubelka-Munk (KM) equations~\cite{kubelka1931article} can be used to reproduce the color of a layer of pigment, the pigment coefficients are difficult to acquire~\cite{okumura2005developing}, so researchers have pursued a simplified model.  
Curtis et al.~\shortcite{Curtis:1997:CW} use a three wavelength model they compute from samples of paint over white and black backgrounds.
In our multispectral scenario given RGB data and a fixed background, direct extraction is ill-posed.
The IMPaSTo system~\cite{Baxter:2004:IMPaSTo} uses a low dimensional approximation of measured pigments to enable realtime rendering.
In contrast, we focus on the problem of decomposing existing artwork.
Xu et al.~\shortcite{xu2007pigment} use a neural network to learn to predict RGB colors from a large number of synthetic examples.  
RealPigment~\cite{Lu:2014:RPC} estimates composited RGB colors from exemplar photos of artist color mixing charts.
In our scenario, we are given the RGB colors and estimate the multispectral scattering and absorption parameters.

\begin{figure}
\centering
\includegraphics[width=0.95\columnwidth]{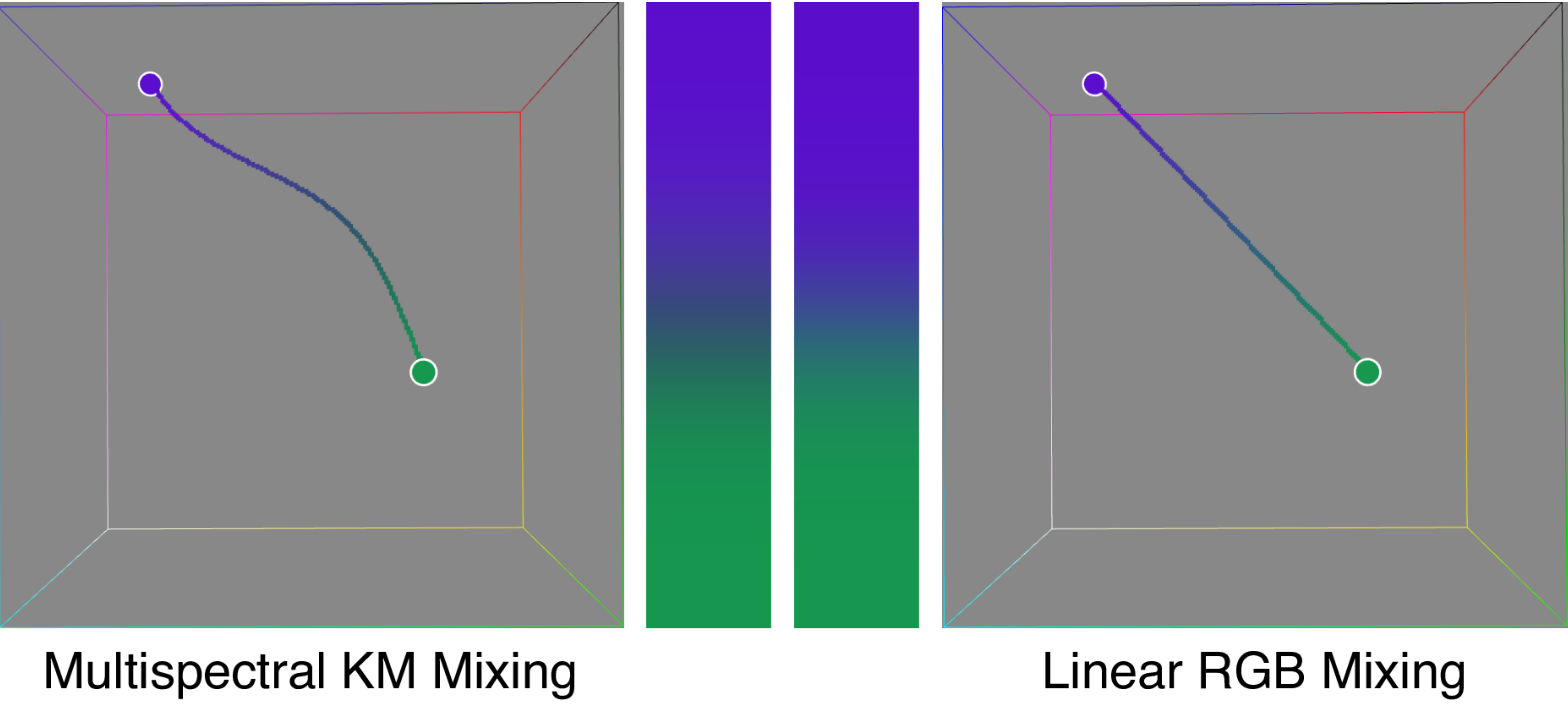}
\caption{Comparing mixing models.  Left: A gradient interpolating between purple and green pigments using the KM equation, and the resulting colors in RGB-space. Right: A gradient interpolating the same pigment colors in RGB-space.}
\label{fig:KM_vs_RGB}
\end{figure}

There is extensive work on multispectral acquisition systems using custom hardware~\cite{boldrini2012hyper}.  Berns et al.~\shortcite{berns2005spectral} use a standard digital camera with a filter array.  Parmar et al.~\shortcite{parmar2012led} use a bank of LED's to capture the scene under different illumination spectra.  Park et al.~\shortcite{park2007multispectral} optimize a set of exposures and LED illuminants to achieve video rates.  Multispectral images have many useful applications.  Ibrahim et al.~\shortcite{Ibrahim2016} demonstrate intrinsic image reconstruction and material identification.  Berns et al.~\shortcite{berns2005comparison} compare a multispectral imager to a point spectrophotometer for measurements of paintings.
Multispectral imaging provides a non-invasive way to preserve paintings and analyze their construction.
Liang et al.~\shortcite{liang2008pigment} used a combination of optical coherence tomography (OCT) imaging with multispectral imaging to identify pigments' reflectance, absorption, and scattering parameters.
Berns et al.~\shortcite{berns2002pigment} estimate the full reflectance spectrum of a painting using a reduced dimension parameterization made from spectra of known KM pigments.  Zhao et al.~\shortcite{zhao2005mapping} achieve better reconstructions by fitting mixtures of known pigments to estimated multispectral reflectances.  
Pelagotti et al.~\shortcite{pelagotti2008multi} and Cosentino~\shortcite{Cosentino2014} both use multispectral images as feature maps to identify single layers of known pigments.  
Most similar to our work, Zhao et al.~\shortcite{zhao2008investigation} use multispectral measurements of Van Gogh's ``Starry Night'' to estimate one parameter masstone KM mixing weights for known pigments to reconstruct the painting.
Delaney et al.~\shortcite{delaney2014use} use fiber optic reflectance spectroscopy and X-ray fluorescence to help identify and map pigments in illuminated manuscripts.
Abed et al.~\shortcite{abed2014pigment} described an approach to identify pigment absorption and scattering parameters and extract pigment concentration maps from a multispectral image via a simplified, one-parameter Kubelka-Munk model.
%
\emph{All} of these works require exotic acquisition hardware, whereas we focus on generating plausible results using standard, easy-to-obtain RGB images. There are plenty of high-quality RGB images of paintings freely available via the Google Art Project.


\section{Theory}
\label{sec:theory}


The intuition behind this work comes from how pigments mix in real versus digital media.  Digital RGB color mixing is a linear operation: all mixtures of two RGB colors lie on the straight line between them in RGB-space.
Mixing two physical pigments with the same two apparent RGB colors, however, produces a curve in RGB-space (Fig.~\ref{fig:KM_vs_RGB}).
The shape of this curve is a function of the multispectral Kubelka-Munk coefficients of the pigments being interpolated.  Our intuition is that those multispectral coefficients can be deduced by the observed shape of a mixing or thinning curve in RGB.

\begin{figure}
	\centering
	\includegraphics[width=0.95\columnwidth]{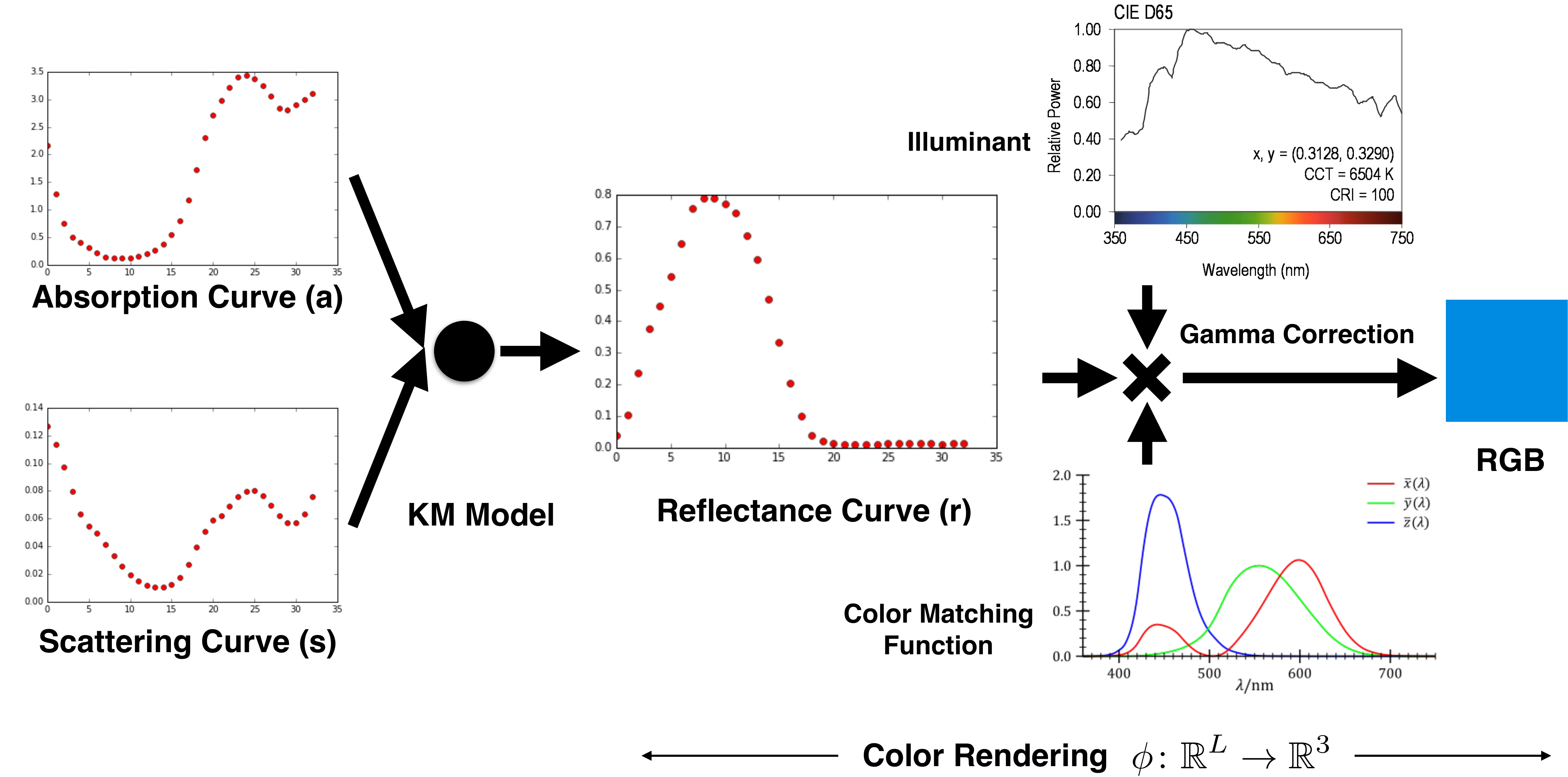}
	\caption{Rendering from multispectral KM coefficients (absorption and scattering) to sRGB color, for cyan pigment, totalling 33 wavelengths ranging from 380nm to 700nm (every 10nm). It is rendered on pure white substrate with pigment thickness equal to 1, under D65 illuminant.}
	\label{fig:km_render}
\end{figure}

\subsection{Kubelka-Munk Equations}
\label{sec:Problem:KM}

The Kubelka-Munk equations (KM) are a physically-inspired model for computing the per-wavelength reflectance value of a layer of homogeneous pigment atop a substrate:
\revtwo{\begin{align}
\label{eq:KM}
\begin{split}
r =\frac{1-\xi(x-y\cdot coth(y s t))} {x-\xi + y\cdot coth(y s t)} \\
x = 1 + \frac{a}{s}, \hspace{0.3cm} y =\sqrt{x^2-1}
\end{split}
\end{align}
}
where $t$ is the thickness of the pigment layer, \revtwo{$a$} and \revtwo{$s$} are the
pigment's absorption and scattering per unit thickness, \revtwo{$\xi$} is the substrate
reflectance, and \revtwo{$r$} is the final reflectance of the pigment layer.  \revtwo{$a$, $s$, $\xi$, and $r$} are all per-wavelength, while the thickness $t$ is constant for all wavelengths.  For convenience, we use $\revtwo{\mathbf{k}}=[\{a_\lambda\},\{s_\lambda\}]$ to represent both KM coefficients with a single \revtwo{vector} variable across all wavelengths $\lambda$.  We denote the vectorized Equation~\ref{eq:KM} as $\revtwo{\mathbf{r} = \mathrm{km}}(\revtwo{\mathbf{k}},\xi,t)$.

Mixtures of pigments are modeled as the weighted average of KM coefficients:
\begin{align}
\revtwo{\mathbf{k}}_{\textit{mix}} = \frac{\sum w_i \revtwo{\mathbf{k}}_i}{\sum w_i}
\end{align}
\revtwo{where $\mathbf{k}_i$ is the $i_{\textit{th}}$ pigment parameter vector.}

To render a KM pigment to RGB requires knowing the pigment's KM coefficients, the substrate reflectance, the layer thickness, the illuminant spectrum, and the color matching functions which map from a reflectance spectrum to a tristimulus value, which can then be converted to linear RGB and gamma corrected to sRGB.  Figure~\ref{fig:km_render} shows the pipeline for a single pigment.  We use the D65 standard illuminant and CIE color matching functions~\cite{ohta2006cie}.

\revtwo{Every pixel has a parameter vector $\mathbf{k}_p$.}
For a pixel $p$ in the image with mixed KM coefficients $\revtwo{\mathbf{k}}_p$, $\revtwo{\mathbf{r}}_p=\revtwo{\mathrm{km}}(\revtwo{\mathbf{k}}_p,\xi,t)$ yields a reflectance spectrum defined at each of the $L$ wavelengths.  We denote the spectrum rendering pipeline in Figure~\ref{fig:km_render} as a function $\phi \colon \mathbb{R}^L \rightarrow \mathbb{R}^3$, so $\revtwo{\phi(\revtwo{\mathbf{r}}_p)}$ is the sRGB color for pixel $p$.
Thus to render an image $\mathbf{I}$ we have:
\begin{align}
\revtwo{\mathbf{I}} = \phi(\revtwo{\mathrm{km}}(\revtwo{\mathbf{K}},\xi,t))
\label{eq:data_equation}
\end{align}
\revtwo{where $\mathbf{K}$ is the matrix of all pixels' pigment parameters.}

In contrast to RGB color compositing, this model is highly non-linear and results in much more of an ``organic'' feel of traditional media paints as compared to digital paintings (Fig.~\ref{fig:KM_vs_RGB}).

\begin{figure}
	\centering
	\includegraphics[width=0.95\columnwidth]{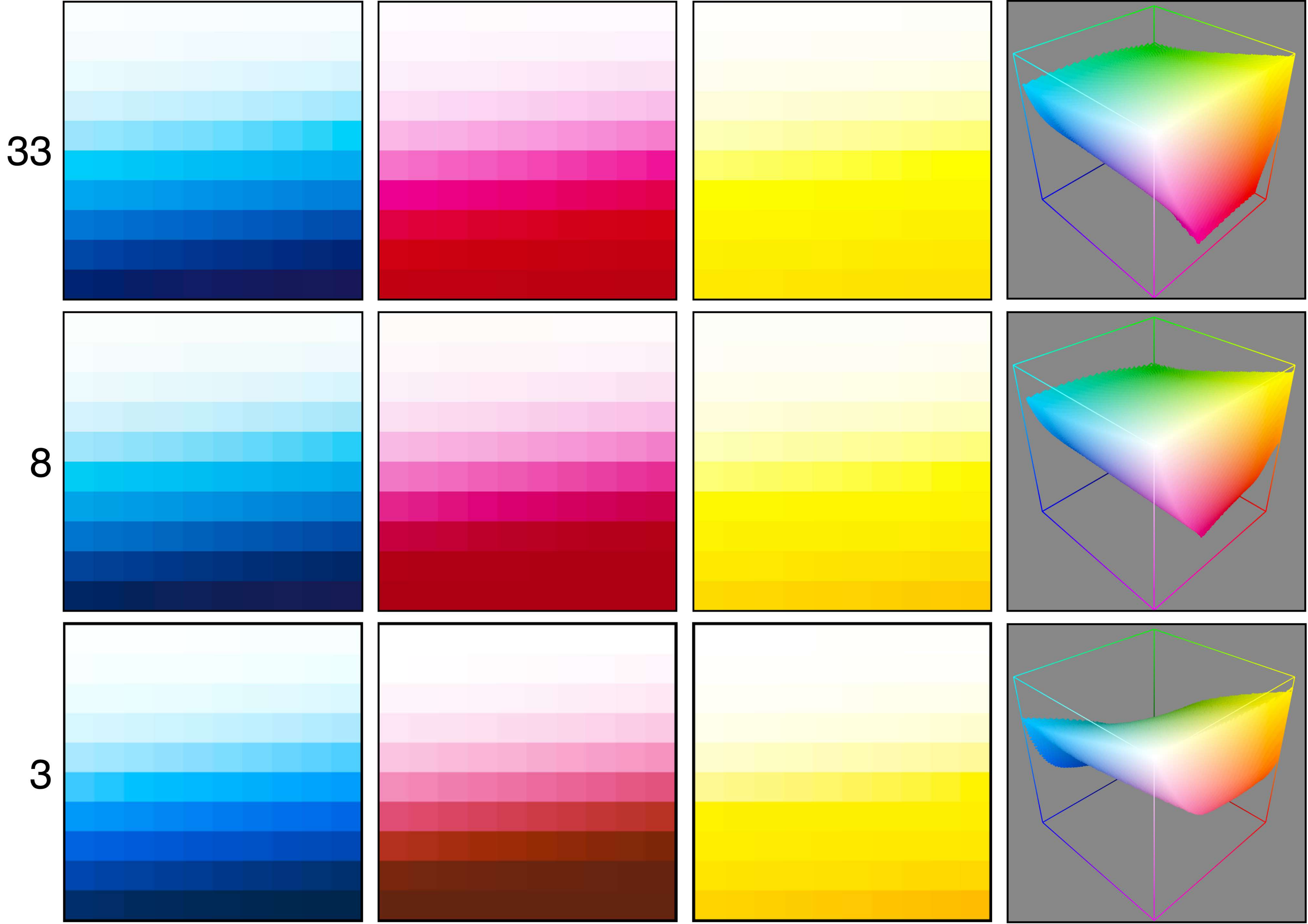}
	\caption{Visualizing rendering with different numbers of wavelengths.  The original cyan, magenta, and yellow pigment coefficients, sampled at 33 wavelengths between 380nm and 700nm, are downsampled to 8 and 3 wavelengths and rendered with varying thickness. The RGB gamuts achieved by mixing them are plotted.  The 8 wavelength gamut appears similar to the 33 wavelength gamut, but the 3-wavelength gamut is significantly distorted.}
	\label{fig:wavelength_influence}
\end{figure}

It is important to consider the required number of wavelengths to simulate.  Too many wavelengths will be difficult to optimize, whereas too few may not be able to accurately reconstruct the image appearance.  We experimented with mixtures of cyan, magenta, and yellow pigments from 33 wavelengths to 3. We found that below 8 wavelengths, the color reproduction loses fidelity (Fig.~\ref{fig:wavelength_influence}).  We can also see that the size of the RGB gamut that can be reconstructed is artificially restricted at 3 wavelengths versus 8.  This is in agreement with prior work such as RealPigment~\cite{Lu:2014:RPC} and IMPaSTo~\cite{Baxter:2004:IMPaSTo}. (A\rev{har}oni-Mack et al.~\shortcite{aharoni2017pigment} is based on a gamut of 3-wavelength KM pigment parameters, which is potentially more restrictive than our gamut of 8-wavelength KM pigment parameters.)

\subsection{Problem Formation}
\label{sec:problem_format}

A painter creates a palette from a set of e.g. tubes of paint, which we call the \emph{primary pigments}.  Every color in the painting is a mixture of these primary pigments.  Therefore, mixtures of the primary pigments' KM coefficients are sufficient to reproduce the RGB color of every pixel in the painting.  Our method estimates the coefficients of a small set of primary pigments to minimize the RGB reconstruction error.

For $L$ wavelengths, each primary pigment $\revtwo{\mathbf{k}}_m$ is a vector of $2L$ coefficients.
We represent the set of $M$ primary pigments as an $M \x 2L$ matrix $\revtwo{\mathbf{H}=[\mathbf{k}_1, \mathbf{k}_2, \hdots \mathbf{k}_M]}^T$.  Every pixel in the painting can be represented as a convex combination of these primary pigments, $\mathbf{w} \cdot \revtwo{\mathbf{H}}$, where \rev{$\mathbf{w}$} is the $1 \x M$ vector of mixing weights (\rev{$\mathbf{0}_{1\x M} \le \mathbf{w} \le \mathbf{1}_{1\x M}$}).
%
We can express all $N$ pixels in the image as the matrix product $\revtwo{\mathbf{K}=\mathbf{WH}}$, where the \rev{$\mathbf{w}$} form the rows of the $N \x M$ matrix $\revtwo{\mathbf{W}}$.
Eq.~\ref{eq:data_equation} becomes:
\begin{align}
\label{eq:data_equation1}
\revtwo{\mathbf{I}}=\phi(\revtwo{\mathrm{km}}(\revtwo{\mathbf{W H}}, \xi, t))
\end{align}
where $\revtwo{\mathbf{I}}$ is the $N \x 3$ matrix of our painting's per-pixel RGB colors.

To simplify the problem, we assume the canvas is pure white ($\revtwo{\xi}=1$).
We further assume that the entire canvas is covered with a single layer of constant thickness $t=1$ paint,
where each pixel's paint is a weighted mixture of pigments.
Thus, our equation becomes:
\begin{align}
\label{eq:data_equation2}
\revtwo{\mathbf{I}}=\phi(\revtwo{\mathrm{km}}(\revtwo{\mathbf{WH}}))
\end{align}
We use Eq.~\ref{eq:data_equation2} to pose an optimization problem:
\begin{align}
\label{eq:whole_optimization}
\begin{split}
E_{\revtwo{\textit{data}}}=\left\| \revtwo{\mathbf{I}}-\phi(\revtwo{\mathrm{km}}(\revtwo{\mathbf{W H}})) \right\|^2 \\
E_{\revtwo{\textit{sum}}}=\left\|\revtwo{\mathbf{W}} \mathbf{1}_{M\x 1} - \mathbf{1}_{N\x 1} \right\|^2 \\
\revtwo{\mathbf{W}}^*,\revtwo{\mathbf{H}}^* = \argmin \{E_{\revtwo{\textit{data}}}+ w_{\revtwo{\textit{sum}}} E_{\revtwo{\textit{sum}}}\}  
\end{split}
\end{align}
where $\mathbf{1}_{M\x 1}$ and $\mathbf{1}_{N\x 1}$ are column vectors of ones, and subject to the constraints \rev{$\mathbf{0}_{N\x M} \leq \revtwo{\mathbf{W}} \leq \mathbf{1}_{N\x M}$ and $\revtwo{\mathbf{H}} > \mathbf{0}_{M\x 2L}$}.
$E_{\revtwo{\textit{sum}}}$ forces our per-pixel weights to sum to one, since each pixel's coefficients are a convex combination of the primary pigments.
As an alternative to $E_{\revtwo{\textit{sum}}}$, one could use $W=\softmax(\revtwo{\mathbf{W}}')$ in $E_{\revtwo{\textit{data}}}$. This would allow unconstrained variation of $\revtwo{\mathbf{W}}'$ while maintaining that the weights (rows) of $\revtwo{\mathbf{W}}$ sum to one.  However, in our experiments we found that $E_{\revtwo{\textit{sum}}}$ has better convergence properties.

\rev{We make the assumption that thickness \revtwo{$t = 1$}, because we are primarily focused on acrylic and oil paints, which are quite thick, especially as compared to watercolor. Our assumption means that we cannot capture impasto effects or thin watercolor effects accurately. Note, however, that the choice of \emph{which} constant thickness value to use is arbitrary. Thickness \rev{$t$} appears in the KM equations as a scale factor for $s$, but neither $a$ nor $s$ appear elsewhere except as a ratio. Therefore, changing the constant thickness \rev{$t$} to another value is equivalent to uniformly scaling all $a$ and $s$.}

\begin{figure}
\centering
\includegraphics[width=0.95\linewidth]{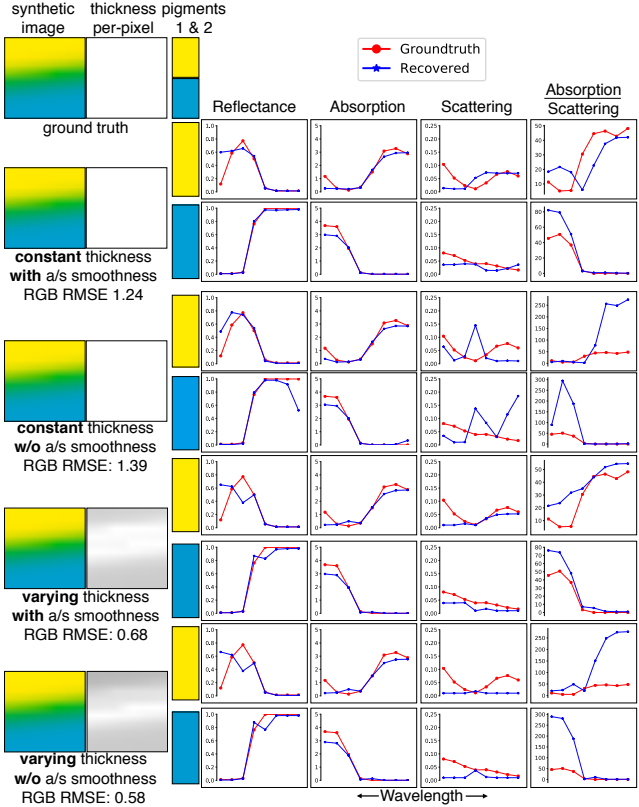}
\caption{\rev{The effects of constant versus varying thickness paint and our $\frac{a}{s}$ masstone smoothness term (Equation~\ref{eq:fixed_W_reg}). The ``synthetic image'' column shows the reconstruction of the ground truth image using two pigments' recovered absorption and scattering parameters shown at right. Allowing paint with varying thickness results in an underconstrained problem. With a smoothness regularization term, the solution deviates from ground truth in our experiments.
The masstone smoothness term results in scattering parameters
that more closely match ground truth.}}
\label{fig:validation}
\end{figure}

\rev{Allowing the thickness to vary introduces an additional degree-of-freedom per pixel. Figure~\ref{fig:validation} shows an experiment in which we solve for two pigments' multispectral $a$ and $s$ parameters and per-pixel mixing weights; we optionally allow
thickness to vary per-pixel. When thickness varies, the problem is under-constrained.
To make the problem tractable, we add a smoothness regularization term. However, this leads to incorrect thickness estimation and less accurate multispectral reflectance (and slower optimization performance).
While varying thickness may be particularly useful for watercolor or translucent paint, we did not pursue it in our thick-paint scenario beyond these initial experiments.}

\subsection{Solution Space}
\label{sec:problem_format:local_minima}

In Eq.~\ref{eq:whole_optimization}, $\revtwo{\mathbf{W}}$ and $\revtwo{\mathbf{H}}$ are both unknown, so we have $NM+2LM$ unknown variables and $3N+N$ known equations, which makes our problem under-constrained for $M>3$.
We can use regularization to make the problem over-constrained. While this results in \emph{a} solution, there are infinitely many solutions to the problem as originally stated for any particular image.  This is for two reasons.

First, $\phi(\cdot)$ projects from $L$-dimensional reflectance spectra to $3$-dimensional \emph{tristimulus} values.
For any given tristimulus value there are infinitely many possible spectra (metamers) that could produce it. This is analogous to seeing only the 2D projection or ``shadow'' of a 3D curve. No matter how many high dimensional samples we obtain, $\phi$ projects them all in parallel.

Second, if there exists $\revtwo{\mathbf{G}}$ s.t.\ $\revtwo{\mathbf{W H}} = \revtwo{\mathbf{W G G}}^{-1} \revtwo{\mathbf{H}}$, for \rev{$\mathbf{0}_{N\x M} \le \revtwo{\mathbf{W G}} \le \mathbf{1}_{N\x M}$, and $\revtwo{\mathbf{G}}^{-1} \revtwo{\mathbf{H}} > \mathbf{0}_{M\x 2L}$}, then $\revtwo{\mathbf{W}}' = \revtwo{\mathbf{W G}}$ and $\revtwo{\mathbf{H}}' = \revtwo{\mathbf{G}}^{-1} \revtwo{\mathbf{H}}$ is another solution that generates the same RGB result.
In a simple geometric sense, $\revtwo{\mathbf{G}}$ could be a rotation or scale of the KM coefficients associated with the primary pigments.  So long as the set of observed pigment parameters all lie within the polytope whose vertices are the rows of $\revtwo{\mathbf{H}}$, then e.g.\ rotations and scales that maintain that property will also produce solutions.  If the colors are near the edges of the gamut, or the pigment parameters are near the edges of the KM space (i.e.\ have small values in the KM coefficients), then there will be very little ``wiggle room'' for the pigments to move.  Conversely, if the set of observed pigment parameters are compact (i.e.\ no KM coefficients near zero), then many different gamuts may be possible.


\section{Method}
\label{sec:method}

Our naively posed optimization problem (Eq.~\ref{eq:whole_optimization})
is too slow to run on an entire, reasonably-sized image at once.
To improve performance, we decompose our task into two subproblems:
estimating primary pigments and estimating per-pixel mixing weights.

\subsection{Estimating Primary Pigments}
\label{sec:method:estimating_primary_pigments}

The first step in our pipeline is to estimate a set of primary pigment coefficients $\revtwo{\mathbf{H}}$ that can reconstruct the painting.  Even for small input images of 0.25 megapixels, doing this estimation over every pixel in the image would be computationally very expensive.
We observe that it is not necessary to consider every pixel, since many pixels contain redundant information.  Therefore, we optimize over a small subset of representative pixels, carefully chosen to well-represent the image's color properties.

To find a small subset of representative pixels $\revtwo{\mathbf{I}}_{\textit{subset}}$, we find the 3D convex hull of the set of RGB colors in the image using the QHull algorithm~\cite{QHull}.  For the images we tested, this usually results in a few hundred unique colors.  These pixels are particularly well suited to the task of estimating the primary pigments because they span the full gamut of the painting's colors. They are guaranteed to include the most extreme combinations of pigments.  Conversely, pixels in the interior of the convex hull are less distinct, resulting in less vibrant recovered primary pigments.

The optimization problem as posed in Eq.~\ref{eq:whole_optimization} is very similar to non-negative matrix factorization---which is non-convex---with the added non-linearities of the KM equation and gamma correction.  Therefore, we use the Alternating Nonlinear Least Squares (ANLS) method for our optimization.

In the first step, we fix the set of primary pigment coefficients $\revtwo{\mathbf{H}}$ and solve for the mixing weights $\revtwo{\mathbf{W}}_{\textit{subset}}$:
\begin{align}
\label{eq:fixed_H_optimization}
\revtwo{\mathbf{W}}_{\textit{subset}}^* = \argmin \{E_{\textit{data}}+ w_{\textit{sum}} E_{\textit{sum}}\}
\end{align}
with the constraint \rev{$\mathbf{0}_{N_1\x M} \leq \revtwo{\mathbf{W}}_{\textit{subset}} \leq \mathbf{1}_{N_1\x M}$, where $N_1$ is the number of representative pixels, and $M$ is the number of pigments}, and $w_{\textit{sum}}=10.0$.

In the second step, we fix $\revtwo{\mathbf{W}}_{\textit{subset}}$ and solve for $\revtwo{\mathbf{H}}$.  When estimating the primary pigments, we add an additional regularization term to avoid creating physically implausible pigment coefficients.  Specifically, KM pigment coefficients $a$ and $s$ should vary smoothly across wavelengths~\cite{Lu:2014:RPC}, and the ratio $\frac{a}{s}$ (which determines the pigment's masstone) should also vary smoothly across wavelengths \rev{(Figure~\ref{fig:validation})}.
We encode these smoothness observations as:
\revtwo{
\begin{align}
\label{eq:fixed_W_reg}
\begin{split}
E_{\textit{smooth}}=\frac{N}{M(L-1)} \sum_{i=1}^M \sum_{\lambda=1}^{L-1} \bigg( w_a & (a_{i,\lambda}-a_{i,\lambda+1})^2 \\
+ w_s & (s_{i,\lambda}-s_{i,\lambda+1})^2 \\
+ w_{\textit{ratio}} \bigg(&\frac{a_{i,\lambda}}{s_{i,\lambda}}-\frac{a_{i,\lambda+1}} {s_{i,\lambda+1}}\bigg)^2 
\bigg)
\end{split}
\end{align}
}
over all $M$ primary pigments and $L$ wavelengths, where $\revtwo{w_a=w_s}=1$ and $\revtwo{w_{\textit{ratio}}}=0.001$ control the relative influence of the terms.
Putting it all together, our optimization for the second step is:
\begin{align}
\label{eq:fixed_W_optimization}
\begin{split}
\revtwo{\mathbf{H}}^* = \argmin \{E_{\textit{data}}+ w_{\textit{smooth}} E_{\textit{smooth}}  \}
\end{split}
\end{align}
with the constraint $\revtwo{\mathbf{H}}>\mathbf{0}_{M\x 2L}$, where $M$ is the number of pigments, and $L$ is the number of wavelengths, and $w_{\textit{smooth}}=0.001$.

\paragraph{Initialization}
As with any non-convex optimization problem, initialization is a key factor in how quickly the solution converges and whether a local minimum is found.  In our case, the solution is not unique, so there are many potential minima to converge upon.  Therefore, initialization is very important for finding a good solution.

\begin{figure*}
\centering
\includegraphics[width=0.95\textwidth]{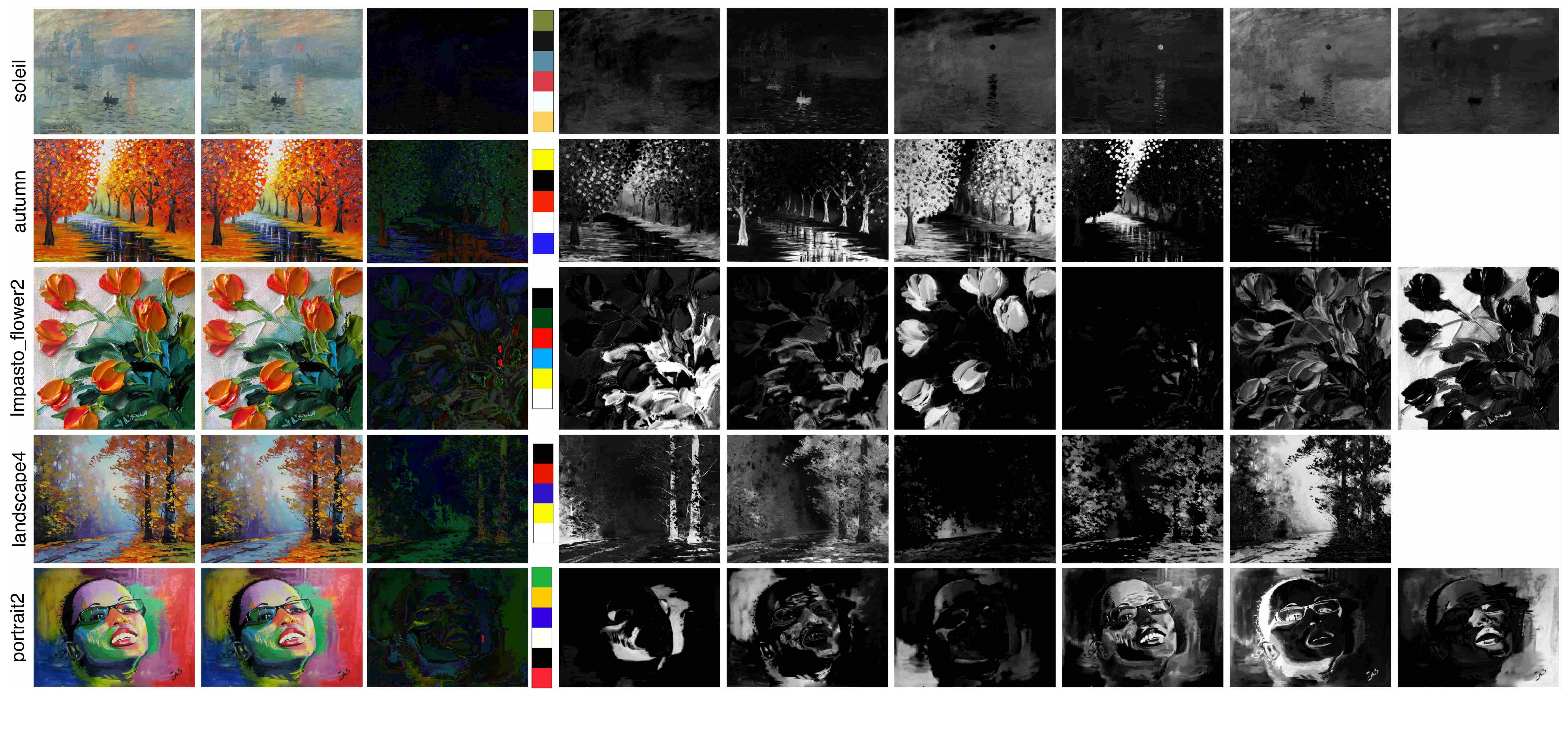}
\caption{Our results for multiple images, one per row.  From left to right the columns are the original image, the reconstruction, the error ($10\x$), the extracted palette, and the mixing weight maps.  Because our pigments are multispectral, we show them as RGB colors rendered on a white canvas with unit thickness. \revthree{From second to end: \copyright \ MontMarteArt, Jan Ironside, Graham Gercken, EAB.PaintingsQatar.}
}
\label{fig:real_images}
\end{figure*}

One option is to initialize randomly, which produces somewhat unpredictable results, though we do find that plausible solutions (where the colors of the primary pigments roughly match a painter's expectations) are often well-represented.  In the absence of other information, we can present the user with the results of e.g.\ ten random initializations to choose their preferred solution.

An alternative is to take advantage of a prior, such as a natural distribution of real pigments in KM space.  We use a set of 26 measured acrylic paints from Okumura~\shortcite{okumura2005developing}.  When the prior is similar to the pigments used in the painting, reconstruction often finds the approximately correct KM coefficients.  When the prior is of a different media than the painting (e.g.\ using an acrylic prior with a watercolor painting), then while the result will have low reconstruction error and look plausible, the mixing properties of the pigments may not be correct (e.g.\ a watercolor painting may have more opaque pigments in the reconstruction than in reality).  When multiple such priors are available, a user could select the correct prior to use for a given painting.  In our experiments, we rely on the dictionary of Okumura's 26 acrylic pigments.  To boost the size of the dictionary, we also include every pair of pigments mixed 50\%, for a total of 351 entries.

To initialize $\revtwo{\mathbf{H}}$ with $M$ pigments using our prior, we start from the convex hull of the RGB colors in the image.  We simplify the convex hull to $M$ vertices as in Tan et al.~\shortcite{Tan:2016:DIL}.  We then match these $M$ RGB colors to the closest matching (Euclidean distance) RGB colors in the dictionary and use the corresponding KM coefficients.  The RGB color of a dictionary pigment is obtained by rendering with the same pipeline as Eq.~\ref{eq:data_equation} (with thickness $t=1$ and substrate $\revtwo{\xi}=1$).
If two convex hull colors match to the same dictionary color, the closer match is used and the other convex hull vertex matches to its second closest dictionary color.

\subsection{Estimating Mixing Weights}
\label{sec:weights}

The second step of our pipeline uses the estimated set of primary pigments to compute per-pixel mixing weights for the entire image.  We use observations about the nature of painting construction to add additional regularization terms, improving convergence and making the results more useful for editing applications.

First, we add a term for per-pixel weights sparsity \rev{\cite{Tan:2016:DIL}}, which encourages each pixel's $M$ pigment weights to be close to 0 or 1:
\begin{align}
E_{\textit{sparse}}= - \frac{1}{M} \left\| \mathbf{1}_{N\x M} - \revtwo{\mathbf{W}} \right\|^2
\end{align}
where $\mathbf{1}_{N\x M}$ is matrix of ones.  This term has the effect of maximizing color separation throughout the painting, so that each pigment influences as small a portion of the image as possible.  This is desirable because it results in more localized pigment editing operations.

Second, we add a term for spatial smoothness of the weights:
\begin{align}
E_{\textit{spatial}}=\frac{1}{M} \left\| \revtwo{\mathbf{S W}} \right\|^2
\end{align}
where $\revtwo{\mathbf{S}}$ is a Laplacian or a bilateral smoothing matrix. We use a bilateral operator~\cite{BarronPoole2016} in order to preserve edges that appear between brush strokes of different colors of paint.  In our experiments, the Laplacian operator blurred edges and caused pigments to incorrectly bleed into image regions.

With these additional terms, our optimization to reconstruct mixing weights becomes:
\begin{align}
\begin{split}
\label{eq:weights_recover_optimize}
\revtwo{\mathbf{W}}_{\textit{all}}^* = \argmin \{E_{\textit{data}} + w_{\textit{sum}} E_{\textit{sum}} & + w_{\textit{sparse}} E_{\textit{sparse}} \\
& + w_{\textit{spatial}} E_{\textit{spatial}} \} 
\end{split}
\end{align}
where $w_{\textit{sum}}=10$, $w_{\textit{sparse}}=0.1$, and $w_{\textit{spatial}}=1$
subject to \rev{$\mathbf{0}_{N\x M} \leq \revtwo{\mathbf{W}}_{\textit{all}} \leq \mathbf{1}_{N\x M}$, where $N$ is the number of pixels in the entire image and $M$ is the number of pigments.}

\begin{algorithm}
\DontPrintSemicolon 
\KwIn{RGB Image $\revtwo{\mathbf{I}}_{N \x 3}$ and user-provided number of primary pigments $M$.}
\KwOut{Primary pigment KM parameters $\revtwo{\mathbf{H}}_{M\x 2L}$ and mixing weights $\revtwo{\mathbf{W}}_{N\x M}$.}

$\revtwo{\mathbf{I}}_{\revtwo{\textit{subset}}}$ $\gets$ vertices of ConvexHull( $\revtwo{\mathbf{I}}$ ) \;
palette$_{\textit{RGB}}$ $\gets$ Simplify( ConvexHull( $\revtwo{\mathbf{I}}$ ), $M$ )\;
$\revtwo{\mathbf{H}}^{(0)} \gets$ ClosestColors( palette$_{\textit{RGB}}$, Okumura mixtures )\;
$i \gets 0$\;
\While{true}{
	$\revtwo{\mathbf{W}}_{\textit{subset}}^{(i)} \gets$ Solve Equation \ref{eq:fixed_H_optimization}( $\revtwo{\mathbf{I}_{\textit{subset}}}$, $\revtwo{\mathbf{H}}^{(i)}$ )\;
	$\revtwo{\mathbf{H}}^{(i+1)} \gets$ Solve Equation \ref{eq:fixed_W_optimization}( $\revtwo{\mathbf{I}}_{\revtwo{\textit{subset}}}$, $\revtwo{\mathbf{W}}_{\textit{subset}}^{(i)}$ )\;
	// Terminate upon small relative change in $\revtwo{\mathbf{H}}$ \;
	// \revtwo{Absolute value and min() are element-wise.} \;
	$\revtwo{\textbf{numerator} \gets | \mathbf{H}^{(i+1)}-\mathbf{H}^{(i)} | }$\;
	$\revtwo{\textbf{denominator} \gets \text{min}( \mathbf{H}^{(i+1)}, \mathbf{H}^{(i)} )}$ \;
	// \revtwo{Maximum is over all elements.} \;
	\If { \revtwo{$\text{maximum}( {\mathbf{numerator}} / {\mathbf{denominator}} ) < 0.001$ \textbf{or} $i=1000$} }
	{$\revtwo{\mathbf{H}}^* \gets \revtwo{\mathbf{H}}^{(i+1)}$\; break\;}
	$i \gets i+1$\;
	}
$\revtwo{\mathbf{W}}_{\textit{all}}^* \gets$ Solve Equation \ref{eq:weights_recover_optimize}( $\revtwo{\mathbf{I}}$, $\revtwo{\mathbf{H}}^*$ )\;
\Return{$\revtwo{\mathbf{H}}^*$, $\revtwo{\mathbf{W}}_{\textit{all}}^*$}\;
\caption{\rev{Extract mixing weights and multi-spectral pigment parameters from single RGB image}}
\label{algo:pipeline}
\end{algorithm}

This optimization is still very large and difficult to solve directly.  Instead, we solve it in a coarse-to-fine manner.  We downsample the image by factors of two until the short edge is less than 80 pixels. We solve the optimization on the smallest image, initializing each pixel's mixing weights to $1/M$. We upsample each solution (mixing weights) as the initialization for the next larger optimization.  We repeat this procedure \rev{to obtain a solution for the original image.}

\begin{figure}
\centering
{\normalsize Primary pigment estimation convergence}\\
\raisebox{1.27in}{\rotatebox[origin=t]{90}{\footnotesize Total energy}}
\includegraphics[width=.95\columnwidth]{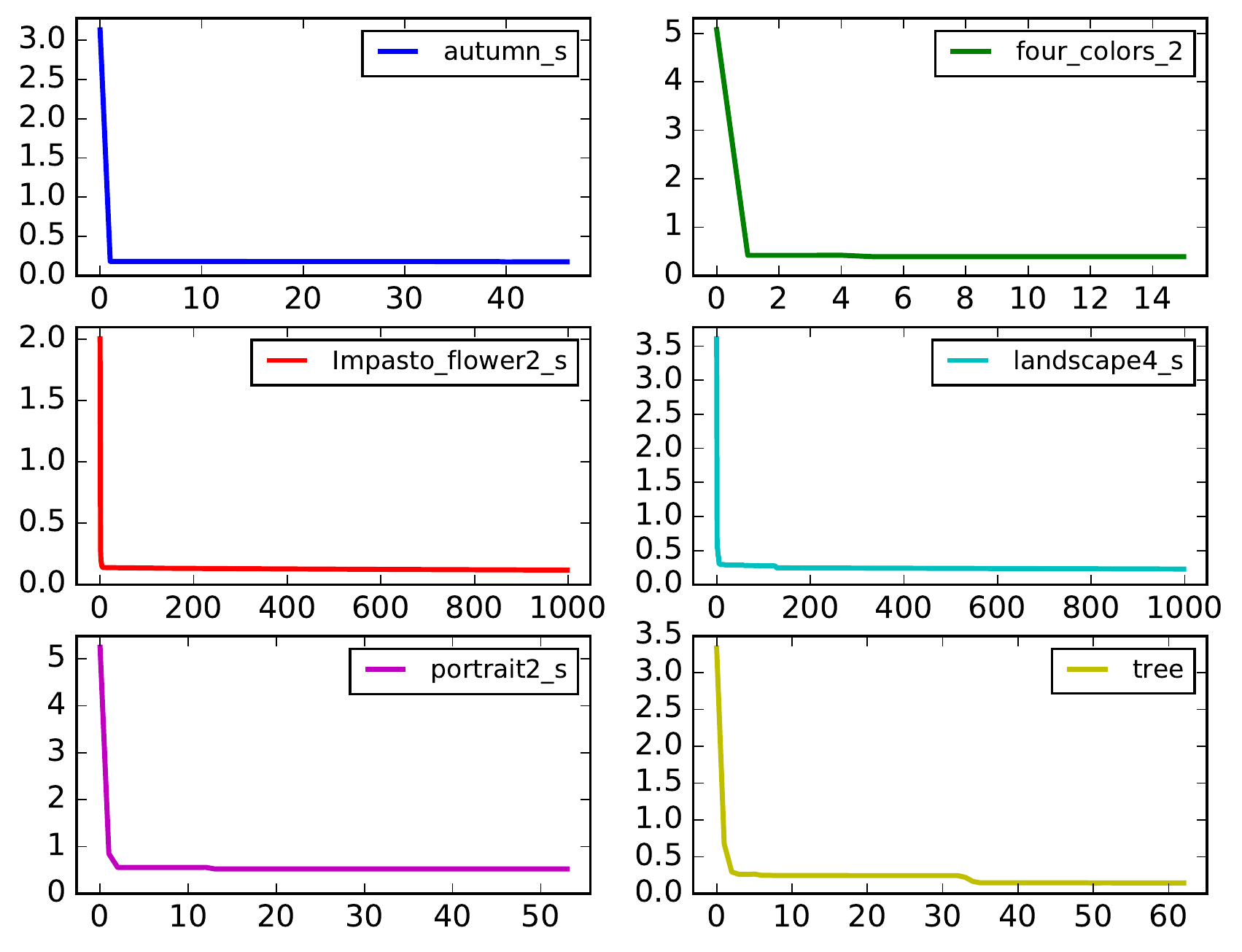}\\
\hspace{2em}{\footnotesize Iteration}
\caption{\revtwo{The total energy for our primary pigment estimation decreases monotonically per iteration ($E_{\textit{data}}+ w_{\textit{smooth}} E_{\textit{smooth}} + w_{\textit{sum}} E_{\textit{sum}}$). Each iteration performs both alternating steps, minimizing Equations~\ref{eq:fixed_H_optimization} and Equation \ref{eq:fixed_W_optimization}.}}
\label{fig:energy_plot}
\end{figure}

\rev{Pseudocode for our method can be found in Algorithm~\ref{algo:pipeline}.
Computational complexity is difficult to analyze because our approach is based on iterative nonlinear optimization
Run-time performance is dominated by the optimization for all pixels' weights (line 15), as discussed in the following section.}
\revtwo{To evaluate the convergence of our pigment parameter estimation's
two alternating optimization steps, we measure the total energy ($E_{\textit{data}}+ w_{\textit{smooth}} E_{\textit{smooth}} + w_{\textit{sum}} E_{\textit{sum}}$) per iteration. Fig.~\ref{fig:energy_plot} plots this for six of the examples used in Table \ref{tbl:perf}.
In all examples, the energy decreases rapidly after a few iterations.
Some examples reach the maximum number of iterations rather than our strict convergence criteria (Algorithm~\ref{algo:pipeline}).}


\section{Results}
\label{sec:results}

To demonstrate our results, we conducted a series of experiments on synthetic and real images, comparing amongst different conditions and with previous work.  All tests were run on a single core of either a 2.53 GHz Intel Xeon E5630 or a 2.50 GHz Intel Core i7-4870HQ, implemented in Python using the \textsc{L-BFGS-B}~\cite{L-BFGS-B} solver.  Runtime information is presented in Table~\ref{tbl:perf}, which shows that we are generally faster than Tan et al.~\shortcite{Tan:2016:DIL}.  Once the primary pigments and mixing weights are estimated, all of our editing applications occur in realtime.


\begin{table}[h]
\small
\centering
\caption{Performance data for Fig.~\protect\ref{fig:real_images} and \protect\ref{fig:compare_to_PD}. Our pipeline extracts $M$ primary pigments in a few seconds and mixing weights maps in less than 10 minutes for a normal size image, with low RGB image reconstruction error.}
\label{tbl:perf}
\resizebox{0.95\columnwidth}{!}{
\begin{tabular}[t]{@{}l c c c c c c@{}}
\toprule
 &  &  & & \textbf{Primary} & \textbf{Weights} & \textbf{RGB} \\
\textbf{Image} & \textbf{Size} & \textbf{M} & \textbf{CPU} & \textbf{(sec)} & \textbf{(sec)} & \textbf{RMSE} \\
\midrule
soleil & $600 \x 467$ & 6 & i7 & 35 & 155 & 0.007 \\
autumn & $600 \x 458$ & 5 & xeon & 16 & 255 & 0.024 \\
four\_colors\_2 & $600 \x 598$ & 4 & i7 & 9 & 211 & 0.020 \\
Impasto\_flower2 & $595 \x 600$ & 6 & xeon & 44 & 615 & 0.02 \\
landscape4 & $600 \x 479$ & 5 & xeon & 26 & 256 & 0.018 \\
portrait2 & $600 \x 441$ & 6 & xeon & 29 & 243 & 0.017 \\
tree & $600 \x 492$ & 4 & i7 & 14 & 151 & 0.016 \\
\bottomrule
\end{tabular}}
\end{table}

\begin{figure}
\centering
\includegraphics[width=0.95\columnwidth]{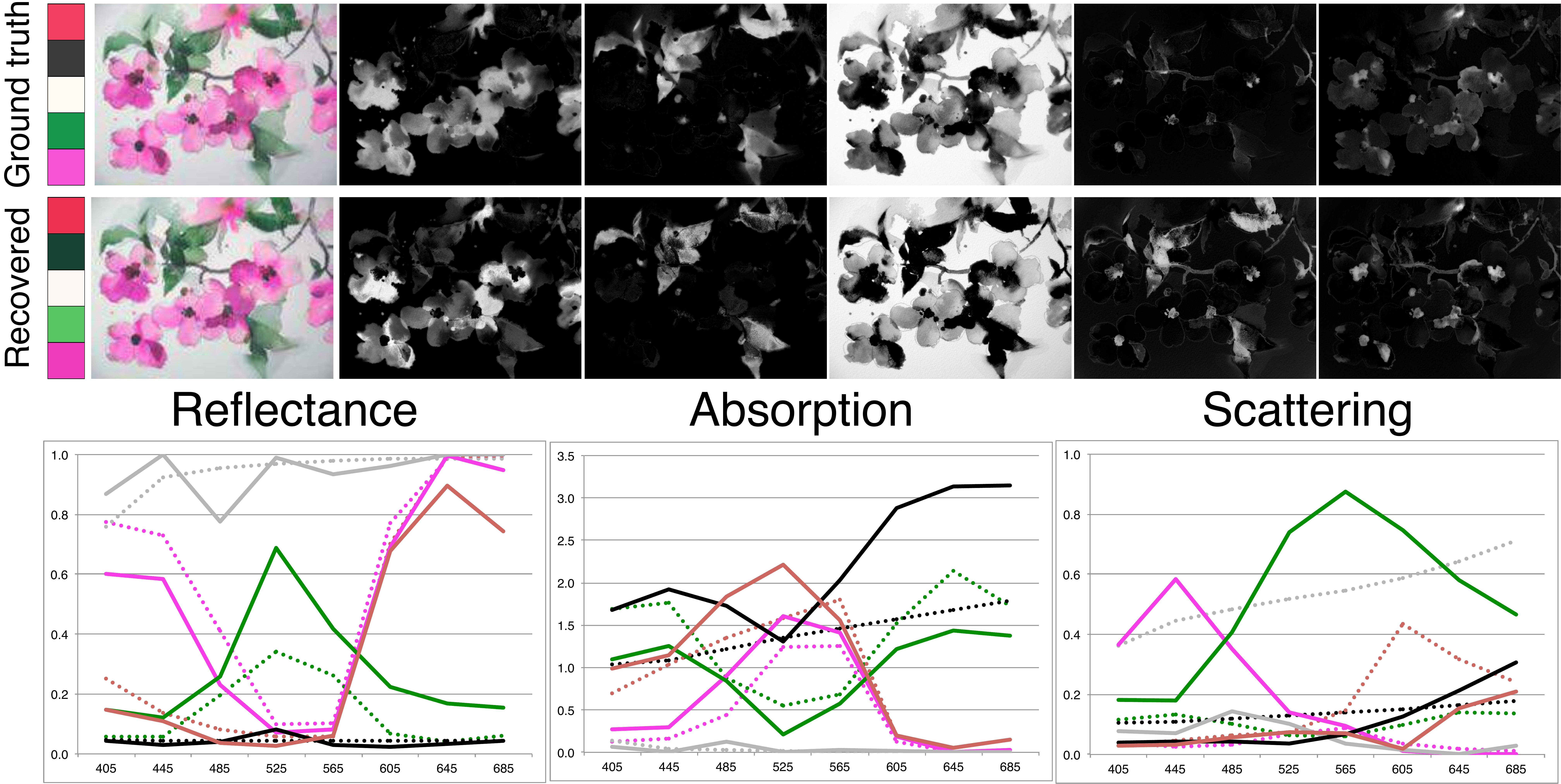}
\caption{Recovering ground truth.  Our reconstruction has low RGB error and the palette and mixing weight maps are similar upon inspection.  The graphs of spectral curves show reflectances are recovered well, but absorption and scattering less so.  Numeric results are in Table~\ref{tbl:error}.  Ground truth curves are dashed, recovered are solid, and colors correspond to palette colors.  Note: ground truth black coefficients are plotted with a scale factor of 0.2 to achieve a similar range as the other pigments. \revthree{\copyright\ Nel Jansen.}}
\label{fig:groundtruth}
\end{figure}

\paragraph{Synthetic Data}
We used synthetic images to evaluate our pipeline's recovery performance against ground truth.  We used our pipeline to obtain weight maps from a painting. We then created five synthetic paintings by randomly choosing sets of pigments from a dataset of measured multi-spectral KM coefficients of real acrylic paint~\cite{okumura2005developing}; mixing them according to our weight maps; and rendering them to sRGB using Eq.~\ref{eq:data_equation}. These five synthetic paintings appear to depict the same flower painted with different colors.  To make our initialization fair, we used a hold-one-out methodology for the pigment dictionary: we removed the five pigments used to construct the synthetic image from the set of candidate pigments for initialization, leaving a dictionary of 21 (plus mixed pigments, so 231 in total).  Fig.~\ref{fig:groundtruth} shows one example of our five synthetic images and its recovery. All reconstruction errors are presented in Table~\ref{tbl:error}.

The results of this experiment are that the pigment coefficients \revtwo{$a$} and \revtwo{$s$} have relatively high error, where for our measured pigments, $\revtwo{a} \in [0,10]$ and $\revtwo{s} \in [0,1]$.  Our reflectance spectra \revtwo{$r$} have lower recover error, $\revtwo{r} \in [0,1]$, because there are many values of \revtwo{$a$} and \revtwo{$s$} that can create the same appearance \revtwo{$r$} (metamers).  Since the pigments are different from ground truth, the recovered weight maps $\revtwo{\mathbf{W}} \in [0,1]$, are different as well. However, the RGB image's reconstruction error is small, generally below the noticeable threshold.  We also tested the weight map recovery step in isolation by using ground truth pigments to estimate $\revtwo{\mathbf{W}}$*, which has a smaller but still significant recovery error. The final RGB* image reconstruction error stays low.  This experiment confirms that there are many solutions to our reconstruction problem, but that we are able to reproduce plausible values.


\begin{table}[h]

\centering
\caption{Reconstruction errors for synthetic data experiments (Fig.~\ref{fig:groundtruth}) with constant weight maps and different pigments.  Each reported number is RMSE, for pigment absorption $\revtwo{a}$ and scattering $\revtwo{s}$ coefficients and reflectance $\revtwo{r}$.  From those pigments, weight map \revtwo{\textbf{W}} and RGB image are recovered.  To test weight map recovery in isolation, \revtwo{\textbf{W}}* and RGB* use the ground truth pigments.
Because there are many solutions (Section~\ref{sec:problem_format:local_minima}),
we cannot recover ground truth parameters (\revtwo{$a$,$s$,\textbf{W}}).
However, the RGB image's reconstruction error is always small and unnoticeable.}
\label{tbl:error}
\begin{tabular}[t]{l c c c c c c c}
\toprule
\textbf{Exp} & \textbf{$a$} & \textbf{$s$} & \textbf{$r$} & \textbf{W} & \textbf{W*} & \textbf{RGB} & \textbf{RGB*} \\
\midrule
1 & 6.1 & 1.2 & 0.3 & 0.114 & 0.060 & 0.019 & 0.023 \\
2 & 1.4 & 0.9 & 0.3 & 0.078 & 0.046 & 0.027 & 0.017 \\
3 & 4.5 & 0.5 & 0.7 & 0.247 & 0.084 & 0.026 & 0.023 \\
4 & 7.1 & 1.2 & 0.6 & 0.166 & 0.055 & 0.033 & 0.024 \\
5 & 1.0 & 0.7 & 0.3 & 0.065 & 0.041 & 0.023 & 0.020 \\
\midrule
Mean & 4.0 & 0.9 & 0.4 & 0.134 & 0.057 & 0.026 & 0.021 \\
Std & 2.7 & 0.3 & 0.2 & 0.074 & 0.017 & 0.005 & 0.003 \\
\bottomrule
\end{tabular}
\end{table}

\begin{figure}
\centering
\includegraphics[width=0.75\columnwidth]{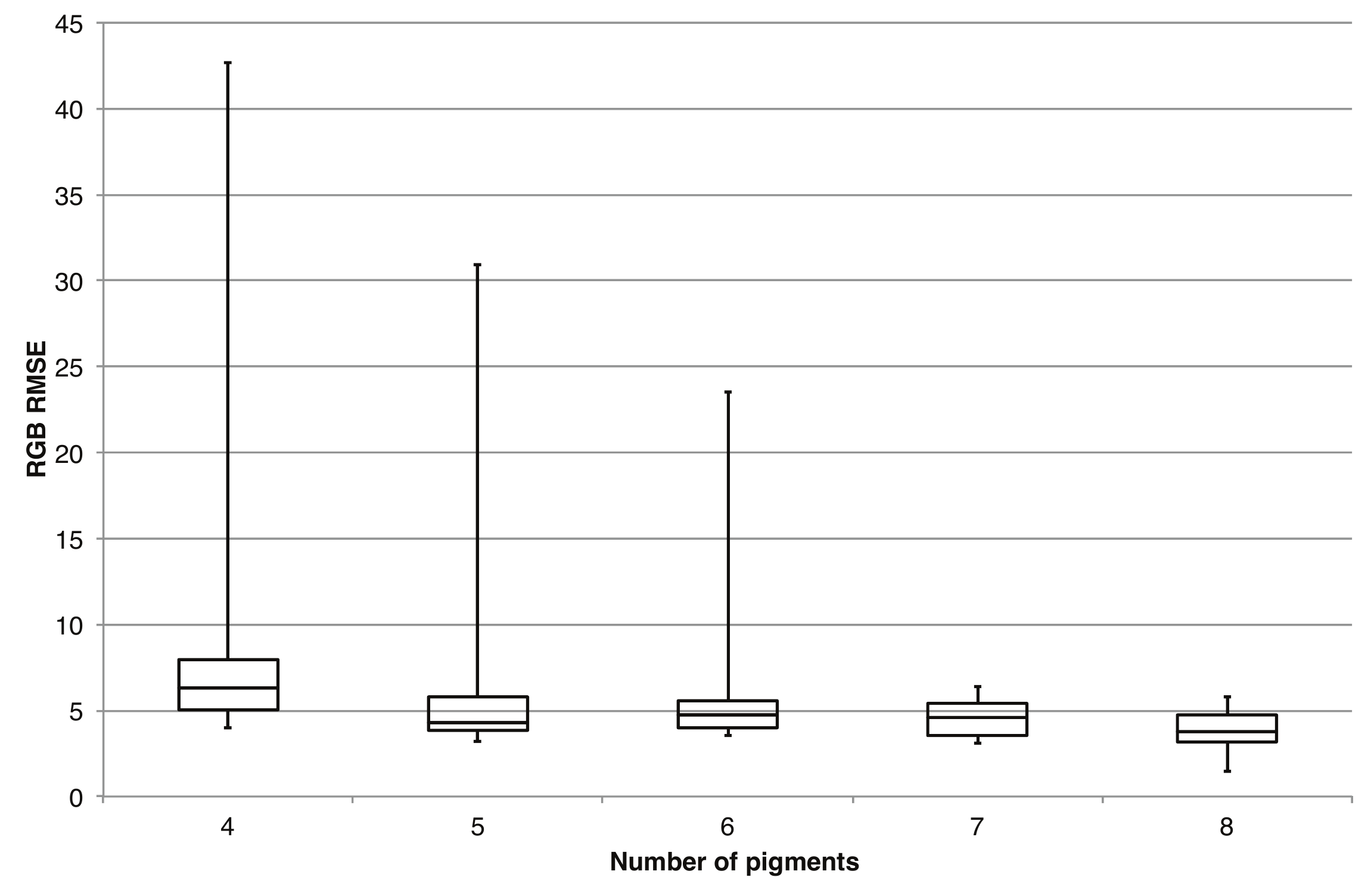}
\caption{ We plot the distribution of RGB RMSE of 12 example images' reconstructions on different palette size. Generally, RMSE will decrease when palette size increase, and RMSE distribution deviation will decrease when palette size increase.}
\label{fig:palette_size}
\end{figure}

\paragraph{Influence of Palette Size}
Since the number of primary pigments is not automatically determined by our algorithm, we evaluated a set of images over a wide range of palette sizes (Fig.~\ref{fig:palette_size}).  Unsurprisingly, as the number of pigments increases, aggregate reconstruction error decreases.  Interestingly, each image seems to have a number of pigments beyond which the RMSE stops decreasing. Intuitively, this would be the natural number of pigments in the painting.  For paintings with very large numbers of pigments, it is unlikely that this property would hold, as eventually a large set of primary pigments would be over-complete and no additional information could be gained.  However, painters often use relatively small palettes in practice.

\begin{figure}
\centering
\includegraphics[width=0.96\columnwidth]{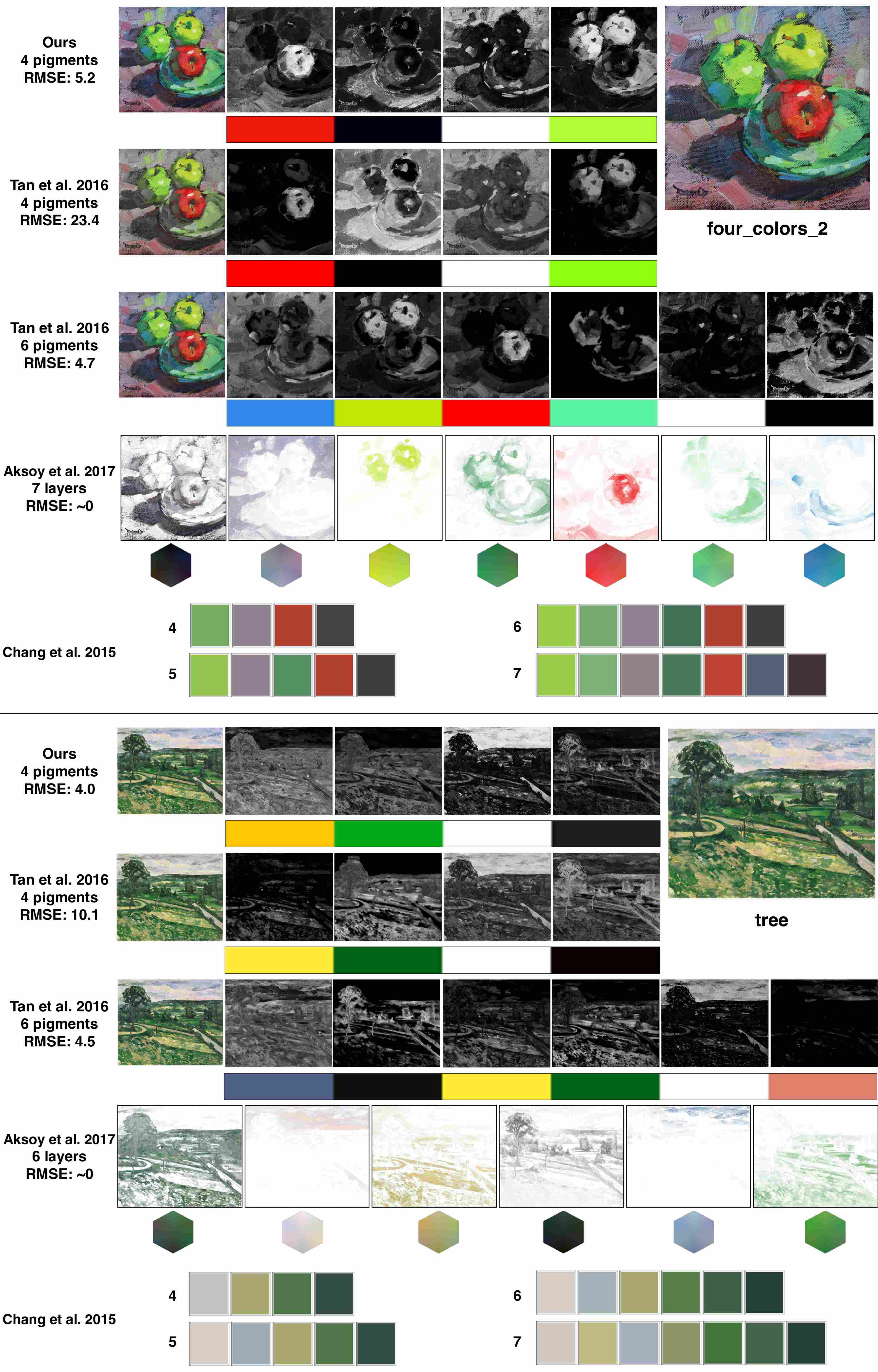}
\caption{Comparison with \rev{the layer decompositions} of Tan et al.~\protect\shortcite{Tan:2016:DIL}
\rev{and Aksoy et al.~\protect\shortcite{aksoy2017unmixing},
and with the palettes extracted by Chang et al.~\protect\shortcite{Chang:2015:PPR}.
The upper four\_colors\_2 example was painted with exactly
four physical pigments.}
When \rev{constrained to four colors}, Tan et al.'s approach has higher reconstruction error.  To match our reconstruction error, Tan et al.'s approach needs to use more colors.
\rev{Aksoy et al.'s approach extracts layers guaranteed to have zero reconstruction error,
but the extracted layers are not composed of a single color.
Chang et al.'s approach extracts a palette whose size is manually chosen by the user.
For the four\_colors\_2 image (top),
Chang et al.'s palettes never contain the known ultramarine blue pigment,
even for very large palettes. 
} \revthree{Top example: \copyright\ Cathleen Rehfeld.}
}
\label{fig:compare_to_PD}
\end{figure}

\begin{figure}
\centering
\includegraphics[width=\columnwidth]{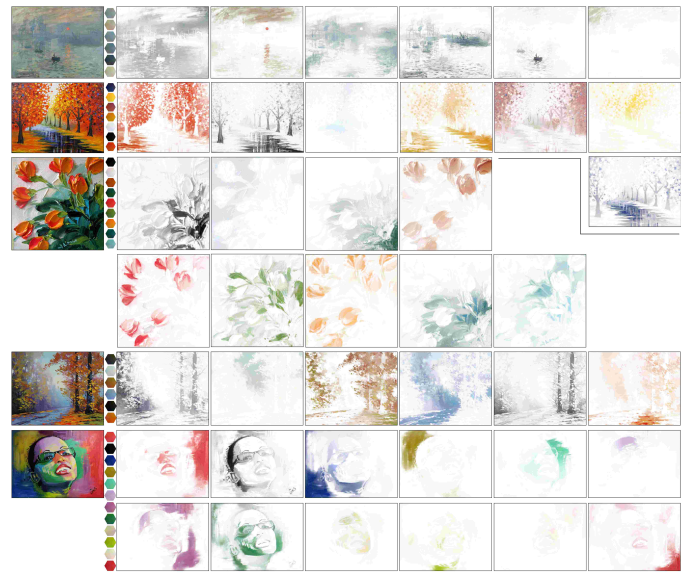}
\caption{
\rev{The approach of Aksoy et al.~\protect\shortcite{aksoy2017unmixing} applied to the same examples as in Figure~\ref{fig:real_images}. The columns show the input image,
their extracted palettes, and their layers. Reconstructions are not shown, because Aksoy et al.'s approach has no reconstruction error. This is because their palettes contain color distributions, not single colors. As a result, their layers are sometimes quite colorful and difficult to edit. The approach automatically chooses a palette size balancing choosing larger (sometimes redundant) palettes with less colorful layers.} \revthree{From second to end: \copyright \ MontMarteArt, Jan Ironside, Graham Gercken, EAB.PaintingsQatar.}
}
\label{fig:Aksoy_results}
\end{figure}

\paragraph{Physical Paintings} 
We show our pipeline running on scans of physical paintings in Figures~\ref{fig:real_images} and \ref{fig:compare_to_PD}: extracted primary pigments, weight maps and reconstructed RGB images.  Reconstruction errors are reported in Table~\ref{tbl:perf}.  We reconstructed with 4 to 6 pigments for every example, but only show the result with the smallest palette that produced low reconstruction error.  Painting four\_colors\_2 is known to have been created with only four paints: titanium white, cadmium yellow lemon, cadmium red, and ultramarine blue.  Our extracted palette's RGB colors are very similar, though the yellow is a bit greenish and the blue is dark.

\paragraph{Comparison to \rev{Tan et al.~\shortcite{Tan:2016:DIL}}.}
Our algorithm uses multispectral pigments with the nonlinear KM model, in contrast to previous work like Tan et al.~\shortcite{Tan:2016:DIL}, which solves a similar problem using a linear compositing RGB model.  Intuitively, we would expect that our model would be able to reconstruct paintings at lower error with fewer parameters. The experiment we show in Fig.~\ref{fig:compare_to_PD} confirms this.  For two paintings, our technique is able to reconstruct the images with low error using only four pigments. Tan et al.~\shortcite{Tan:2016:DIL}'s algorithm results in much higher error for the same number of colors.  In order to achieve a similar RGB reconstruction error, Tan et al.~\shortcite{Tan:2016:DIL} must increase the palette to six colors.
In general, it is easier to edit a painting with a smaller palette.

\rev{\paragraph{Comparison to Aksoy et al.~\shortcite{aksoy2017unmixing}.}
Figures~\ref{fig:compare_to_PD} and \ref{fig:Aksoy_results} show the same examples
decomposed using the approach of Aksoy et al.~\shortcite{aksoy2017unmixing}.
Their approach extracts additive linear RGB mixing layers.
The layers contain color distributions, not single colors,
though their approach guarantees zero reconstruction error.
The number of layers is automatically selected, balancing the colorfulness of layers against smaller palettes.
The colorful layers are difficult to edit, since color distributions must be modified.}

\rev{\paragraph{Comparison to Chang et al.~\shortcite{Chang:2015:PPR}.}
Figure~\ref{fig:compare_to_PD} shows palettes for the same input images
extracted by the approach of Chang et al.~\shortcite{Chang:2015:PPR}.
In their approach, the palette size is manually chosen by the user.
For the four\_colors\_2 image with known four ground truth pigments,
even very large palettes never contains a color similar to ultramarine blue.
Instead, ``redundant'' colors are added.}

\begin{figure}
\centering
\includegraphics[width=\columnwidth]{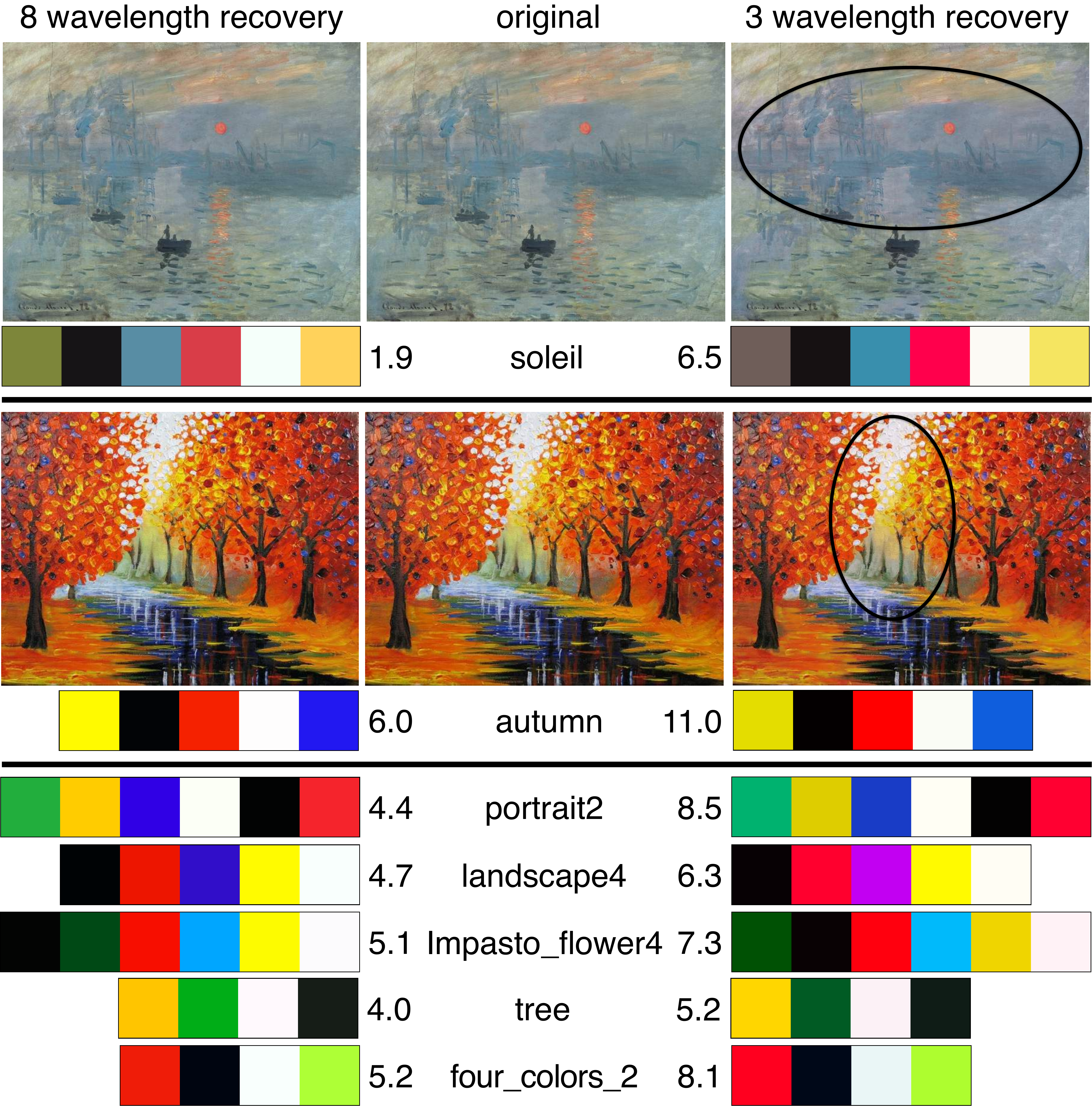}
\caption{Comparison of 3 and 8 wavelength recovery, with RGB RMSE. We find 3 wavelength reconstruction error is higher for all examples. Soleil and autumn example show color distortion. \revthree{Second example: \copyright\ MontMarteArt.}}
\label{fig:3_wavelength}
\end{figure}

\paragraph{Influence of Wavelengths}
Our pipeline recovers 8 wavelength pigment absorption and scattering coefficients, because of the experiment in Figure~\ref{fig:wavelength_influence} that shows a limited RGB gamut for 3 wavelength rendering.  For completeness, we compare with 3 wavelength recovery in Figure~\ref{fig:3_wavelength}.  As 3 wavelength recovery is not multispectral anymore, we slightly amend our model equation, Eq.~\ref{eq:data_equation}: we change the illuminant from D65 to pure white ($\mathbf{1}_{3\x 1}$), and we set the color matching function to be the identity matrix ($\revtwo{\mathbf{I}}_{3\x 3}$).  This has the effect of directly mapping the \revtwo{$a$} and \revtwo{$s$} coefficients to the RGB color channels, as done by Curtis et al.~\shortcite{Curtis:1997:CW}.

We find that 3 wavelength recovery has larger RGB reconstruction RMSE for the same size palettes in all of our experiments, though many of the achieved errors are still low enough to be generally unnoticeable.  For some images, such as the two pictured in Fig.~\ref{fig:3_wavelength}, there is obvious color distortion.  We believe this is due to the restricted gamut of the 3 wavelength pigment model, which has a significant (visible) impact only on paintings that include colors in those extreme portions of the gamut, notably certain greens and reds.  For paintings with colors entirely within the 3 wavelength gamut, the differences will be negligible.

We also compare 3 wavelength recovery with linear RGB of Tan et al.~\shortcite{Tan:2016:DIL} on example images tree and four\_colors\_2 in Fig.~\ref{fig:compare_to_PD} and \ref{fig:3_wavelength}.  The 3 wavelength KM recovery still produces better RGB reconstruction error for the same number of colors than the linear model.



\section{Applications}
\label{sec:applications}

Once we have analyzed a painting to extract its primary pigments (inset for most figures) and mixing weight maps, we can re-pose a number of image editing operations in pigment space to enable interesting paint-aware applications.

\begin{figure}
	\centering
	\includegraphics[width=0.92\columnwidth]{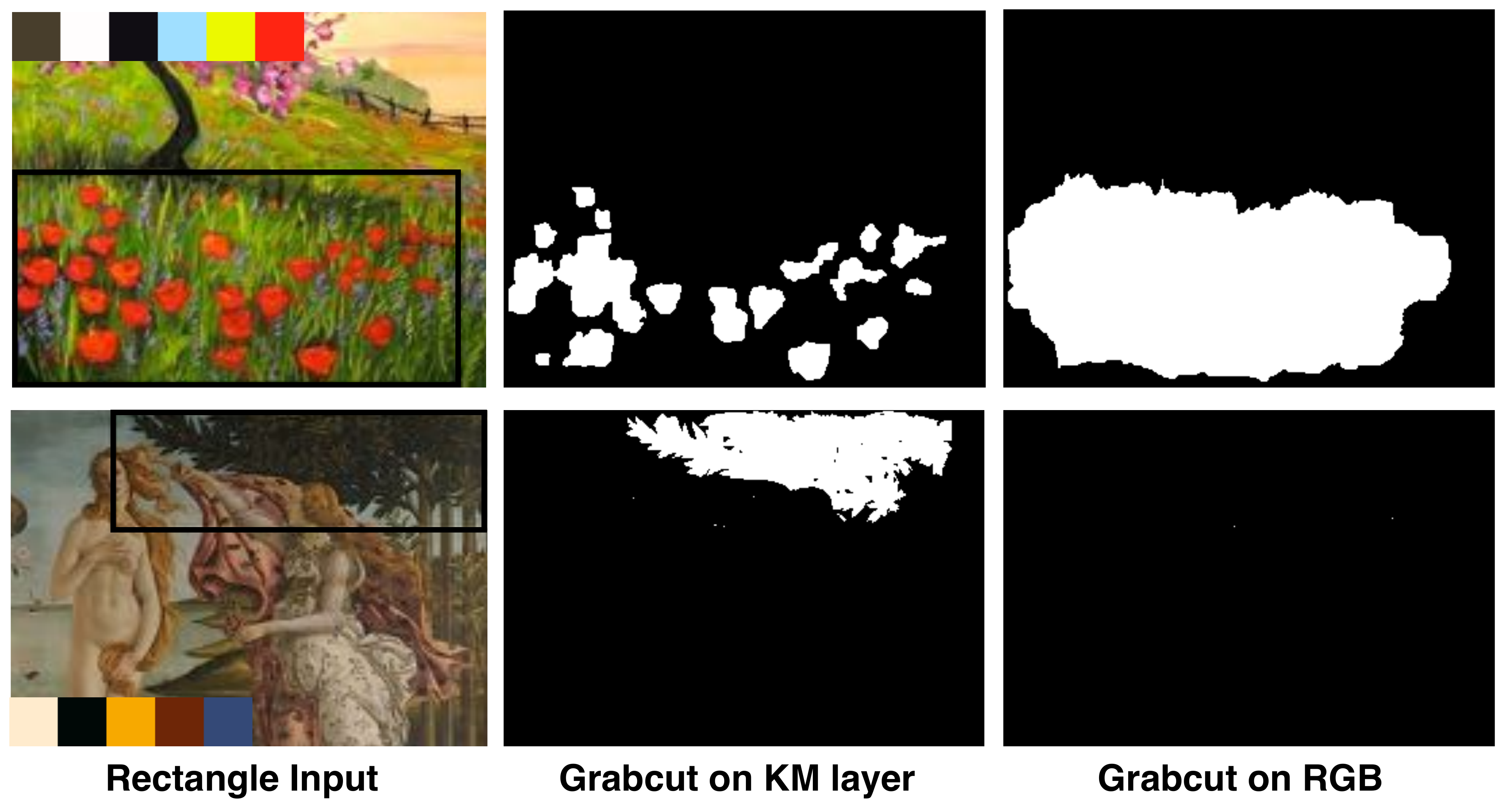}
	\caption{GrabCut on selected KM pigment mixing weights (top: red, bottom: black) outperforms GrabCut on the RGB image. \revthree{First example: \copyright\ Patty Baker.}}
	\label{fig:GrabcutResults}
\end{figure}

\subsection{Masking}

Selection masking in images of paintings can be improved by optimizing on pigment weights instead of RGB colors.  Semantic image boundaries are likely to correspond with changes in paint, whereas RGB edges may be less obvious, when different paint mixtures are used to paint distinct objects.  Also, paint thickness can create lighting variations across the surface of the painting that can confuse RGB boundary analysis.  We demonstrate a standard GrabCut~\cite{Rother:2004:GIF} implementation on two paintings on pigment maps vs. RGB values in Figure~\ref{fig:GrabcutResults}, which clearly shows improved localization of painted objects in the black rectangular regions. GrabCut was performed on the red pigment for the top painting, and on the black pigment for the bottom painting. No background and foreground scribbles are provided to the GrabCut algorithm.

\begin{figure}
	\centering
	\includegraphics[width=0.92\columnwidth]{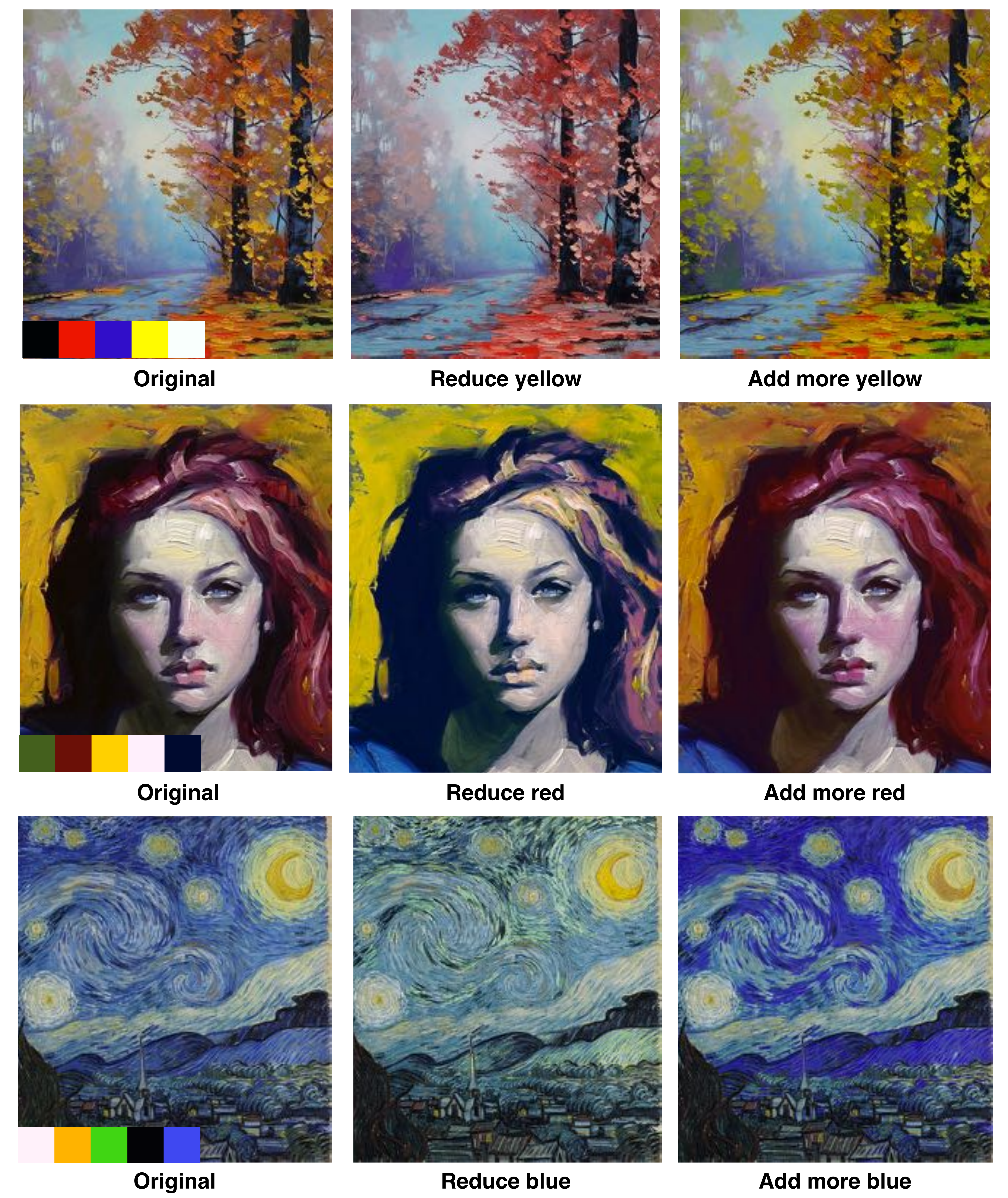}
	\caption{Adjusting the absolute mixing weight of a pigment without re-normalizing the weight sum creates variations that would be difficult to reproduce in RGB. \revthree{First and second: \copyright\  Graham Gercken, John Larriva.}
}
	\label{fig:MixingWeightAdjustment}
\end{figure}

\subsection{Adjustments}

The pigment mixing maps provide a novel parameterization for image edits that may be difficult in RGB, by adjusting the relative values of the mixing weights, or adjusting the coefficients of the extracted pigments.  First, we can vary the weight of any of the extracted pigments by scaling its map up or down and optionally re-normalizing the per-pixel weight sum to one.  For a painting that has e.g. yellow pigment, this change corresponds to varying the amount of yellow in the image, in a way that would be difficult to reproduce using the features of a digital image manipulation program (Fig.~\ref{fig:MixingWeightAdjustment}). 

\begin{figure}
	\centering
	\subfloat[Adjusting the absolute mixing weights of black and white pigments. The effects (middle two images) are different from adjusting the brightness level using photo manipulation software (rightmost image). Increasing the mixing weights of all layers (bottom left image) results in pigments reaching their respective masstones.\revthree{\copyright\ Pamela Gatens}]{%
		\includegraphics[width=0.92\columnwidth]{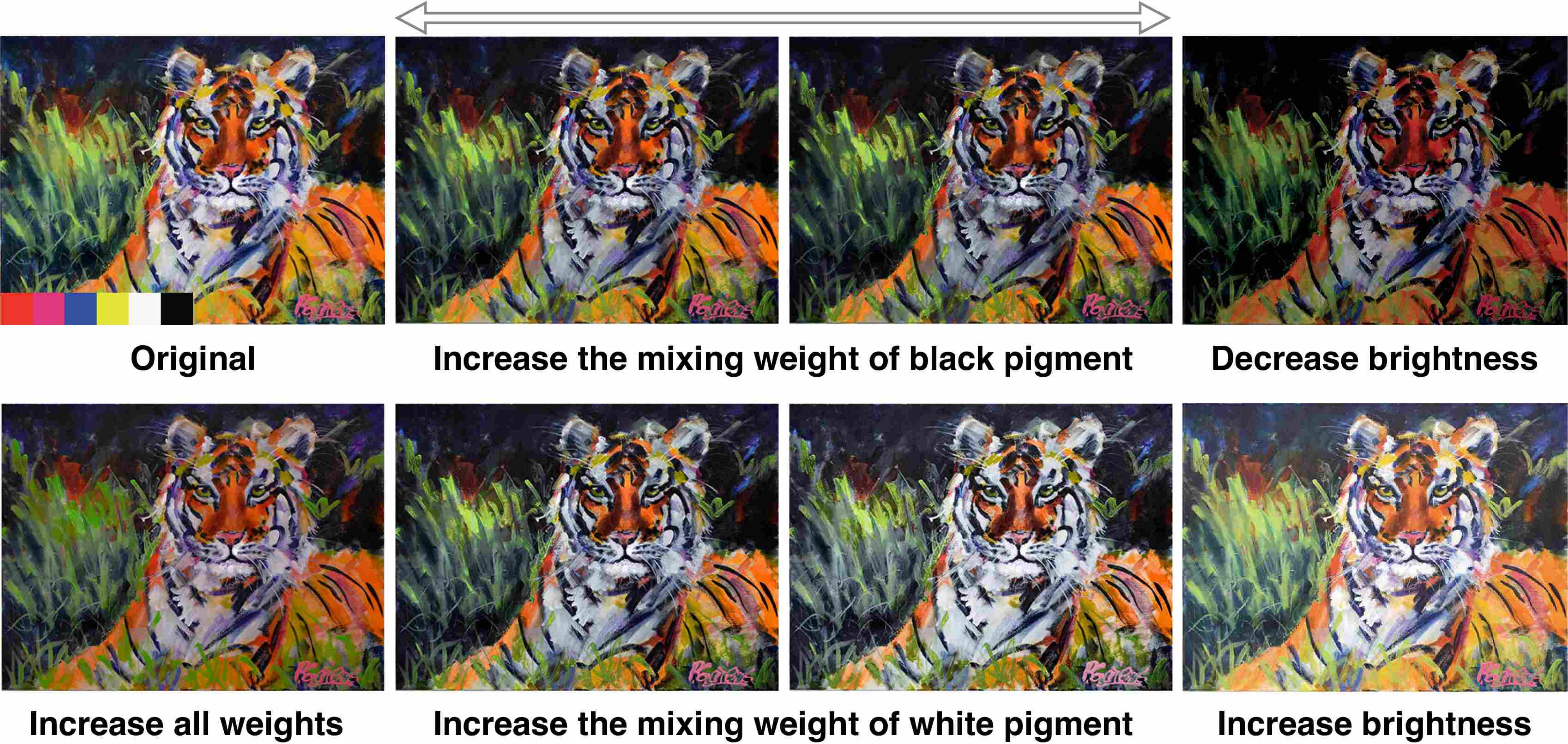}
		\label{fig:MixingWeightAsBrightnessAdjustment}
	}\par
	\subfloat[Adjusting the scattering cofficients of the green pigment. Changing the scattering coefficients produces different effect from manipulating the mixing weights (rightmost image). \revthree{\copyright \ Mark Adam Webster}]{%
		\includegraphics[width=0.92\columnwidth]{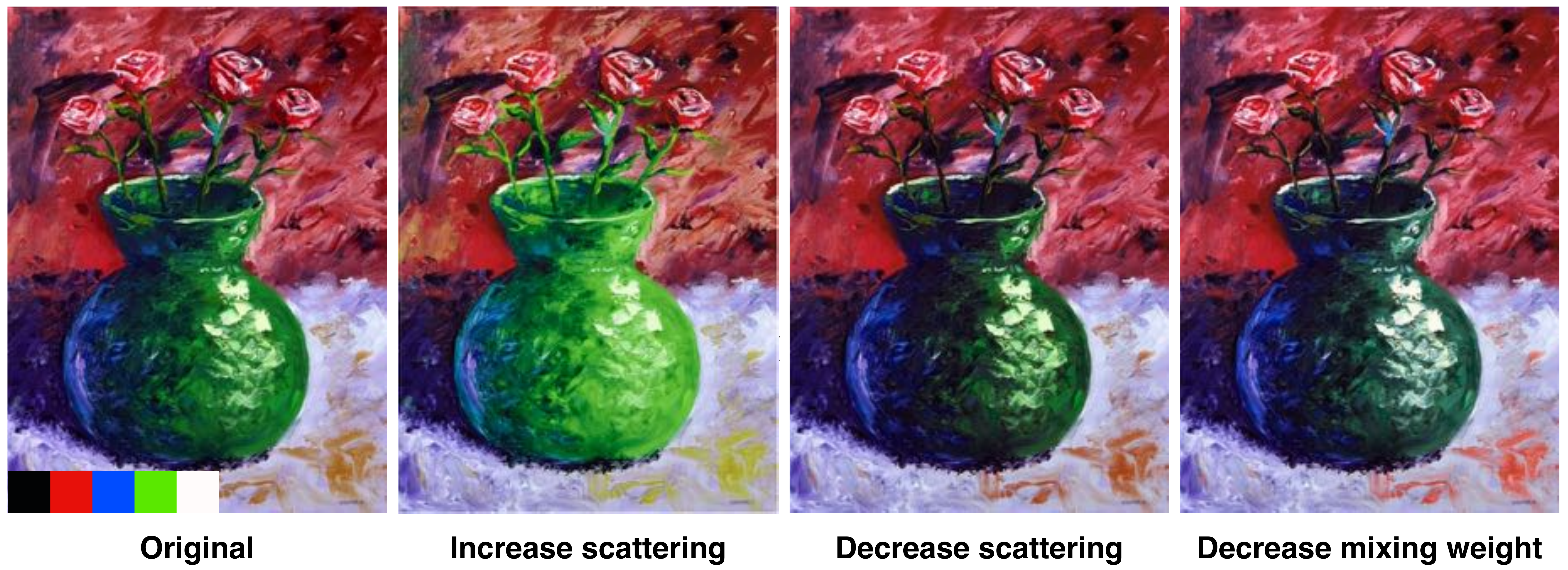}
		\label{fig:ScatteringAdjustment}
	}
	\caption{Tonal adjustments in pigment space.}
	\label{fig:AdjustmentResults}
\end{figure}

Similarly, most extracted palettes include some white and black pigments for creating tints and shades.  Adjusting the relative weights of these pigments is akin to adjusting the brightness and contrast of an image, but again with different results.  For example, in Fig.~\ref{fig:MixingWeightAsBrightnessAdjustment}, the result of increasing the black weight is more like emphasizing shadows and detail, instead of just darkening, while the result of increasing the white weight is desaturation of the colors.

The KM coefficients of the pigments can also be relatively adjusted for interesting effects.  Fig.~\ref{fig:ScatteringAdjustment} shows scaling the per-wavelength scattering coefficients of the green pigment, while keeping absorptions constant.  Increasing scattering means that more light will be reflected back, so in some sense this is similar to brightening the green and making it more opaque, while decreasing scattering creates a darker green that absorbs more than it scatters, so perhaps more like a stained glass. Changing the scattering coefficients produces different hues of green compared to manipulating the pigment mixing weight (rightmost image in Fig.~\ref{fig:ScatteringAdjustment}).

\subsection{Recoloring}

\begin{figure}[t!]
	\centering
	\subfloat[We use our palette's RGB colors for layers in Tan et al.~\protect\shortcite{Tan:2016:DIL} for direct comparison. Top: blue is replaced by green.  Middle: red is replaced by blue.  Bottom: yellow is replaced with a different yellow.\label{fig:RecoloringResults} \revthree{First and second: \copyright\ Pamela Gatens, Patti Mollica.}]{
		\includegraphics[width=0.92\columnwidth]{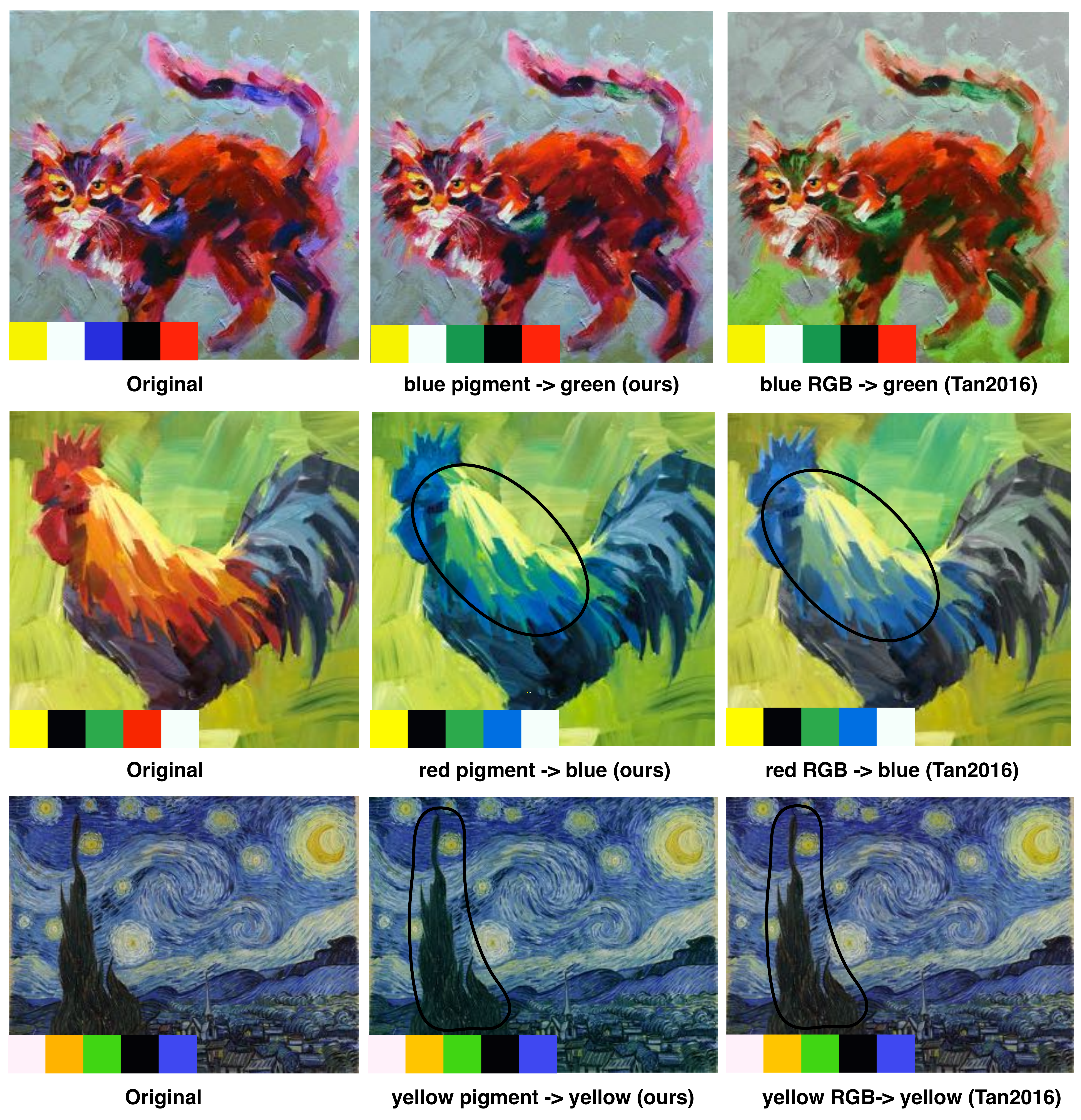}
	} \par
	\subfloat[Each method extracts its own palette from the input image, so we attempt to mimic our result as closely as possible.  Tan et al.~\protect\shortcite{Tan:2016:DIL} suffers from lack of sparsity, while Chang et al.~\protect\shortcite{Chang:2015:PPR} has surprising local colors (red arrows). \label{fig:global_recoloring_comparison} \revthree{\copyright\ Jan Ironside.} ]{
		\includegraphics[width=0.92\columnwidth]{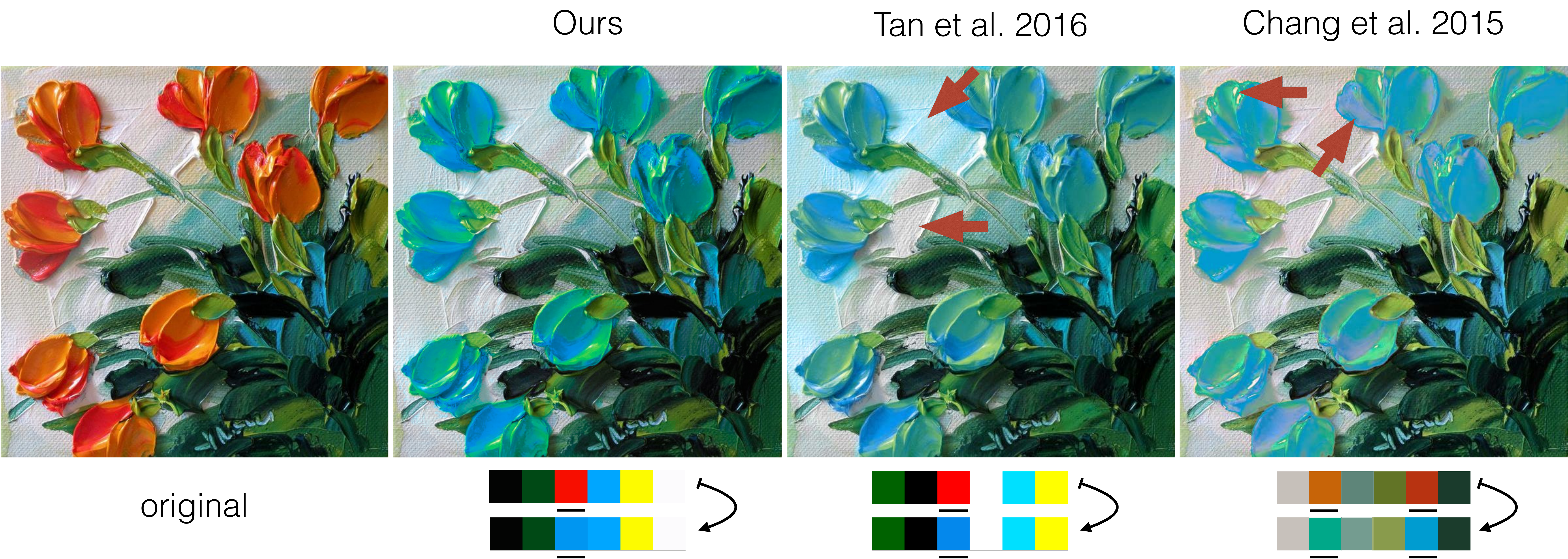}
	}
	\caption{Recoloring comparisons.}
	\label{fig:recoloring}
\end{figure}

Previous work focused on recoloring images by changing the extracted palette colors.
Tan et al.~\shortcite{Tan:2016:DIL} reconstructs each image pixel as a set of RGB layers, so changing a palette color has a straightforward impact on the resulting image.  Our recoloring result is similar to Tan et al.~\shortcite{Tan:2016:DIL}, with the difference being that we replace KM pigments in the extracted palette with other KM pigments (from Okumura~\shortcite{okumura2005developing}) and re-render the image, creating different mixed colors in the style of real traditional media paints. Fig.~\ref{fig:RecoloringResults} shows three examples.  To enable a more direct comparison, we use our extracted palette RGB colors as the layer colors in Tan et al.~\shortcite{Tan:2016:DIL}.
In the cat painting, the KM mixing weight map for the blue pigment is sparse and therefore the recoloring effect is localized on the body of the cat. The weight map from Tan et al.~\shortcite{Tan:2016:DIL} has non-zero values in the background resulting in recoloring artifacts. For the rooster painting, using our KM model, more vibrant green is obtained from mixing yellow and blue in the circled region. For Starry Night, when swapping the extracted yellow pigment with a different yellow, the KM recoloring result reveals the green hue in the new yellow pigment, whereas the RGB recoloring result is similar to the original painting since the two yellow pigments have similar masstones in RGB space.
Fig.~\ref{fig:global_recoloring_comparison} shows a different recoloring comparison between Tan et al.~\shortcite{Tan:2016:DIL}, Chang et al.~\shortcite{Chang:2015:PPR} and ours. Both Tan et al.~\shortcite{Tan:2016:DIL} and Chang et al.~\shortcite{Chang:2015:PPR} have color artifacts when using their own pipelines to recolor the painting to be similar to our result. 

\subsection{Cut, Copy, Paste}

\begin{figure}[t!]
	\centering
	\includegraphics[width=0.92\columnwidth]{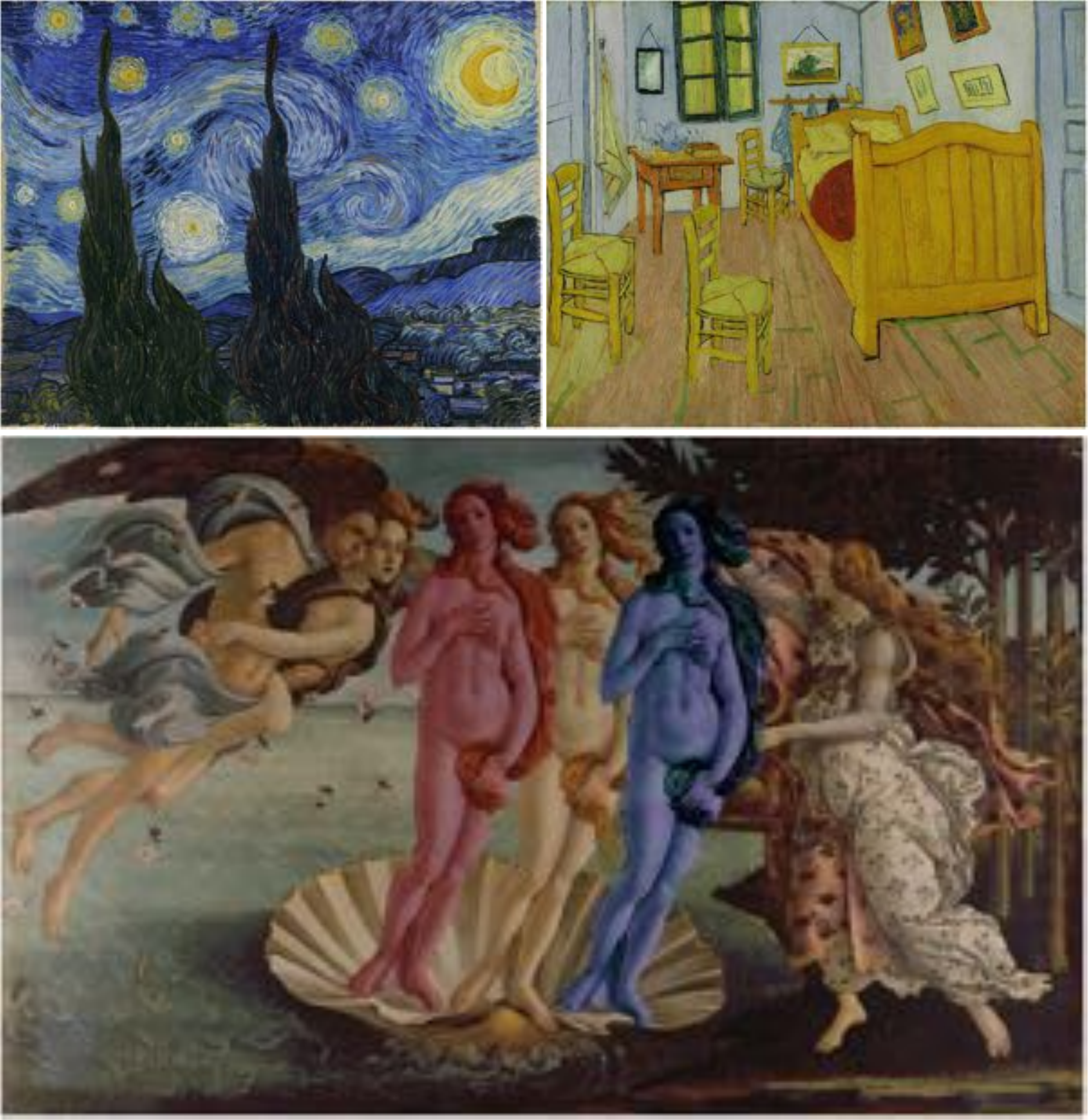}
	\caption{Results of copy paste in pigment space.  Each classical painting has been modified by selecting some set of pigments from a region of pixels, and adding them as a new layer on top elsewhere in the image.  While the pasted regions are not identical to the copied regions (as they would be with standard RGB copy paste), they appear as if they were painted as part of the image.}
	\label{fig:CopyPasteResults}
\end{figure}

From a selection mask, we can use the pigment weight maps to do painterly cut, copy, and paste operations on images as well.  For copy-paste, the user can specify a mask (using any mechanism) and the subset of pigments to copy.  The selected region can then be pasted elsewhere as a new layer of paint on top of the image and re-composited.  The paste operation can adjust paint properties simultaneously such as the thickness of the pasted layer, to achieve different compositing results, or can be added into the mixture model as additional paint mixed into the painting layer, with relative scaling and renormalization.  These options result in different painterly variations on standard image copy-pasting (Fig.~\ref{fig:CopyPasteResults} and Fig.~\ref{fig:teaser}).

The cut operation deletes the selected pixels' pigments from the painting, for which inpainting fills the resulting hole (Fig.~\ref{fig:teaser}).  We use a fast marching method~\cite{telea2004image}, though alternatives such as PatchMatch~\cite{Barnes:2009:PAR} would also work, so long as they can operate on arbitrary numbers of image channels.

\subsection{Palette Summarization}

\begin{figure}[t!]
	\centering
	\subfloat[Palette summarization applied to photos, as compared to Tan et al.~\protect\shortcite{Tan:2016:DIL}, Color, and Chang et al.~\protect\shortcite{Chang:2015:PPR}.]{%
		\includegraphics[width=0.45\columnwidth]{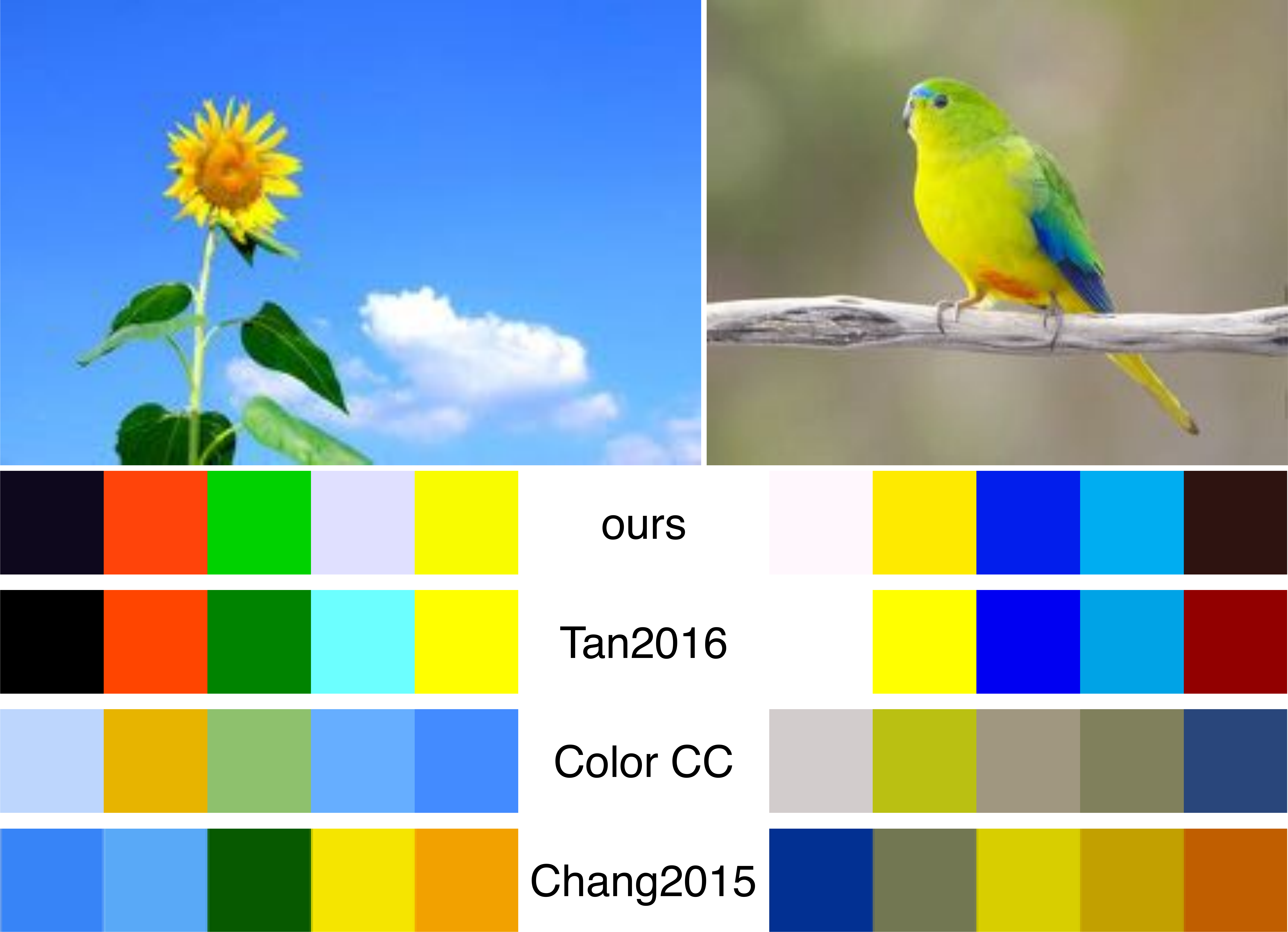}%
		\label{fig:PhotoPaletteResults}
	}\hspace{0.05\columnwidth}
	\subfloat[Summarizations of Van Gogh's paintings arranged by year to show evolution of style.]{%
		\includegraphics[width=0.44\columnwidth]{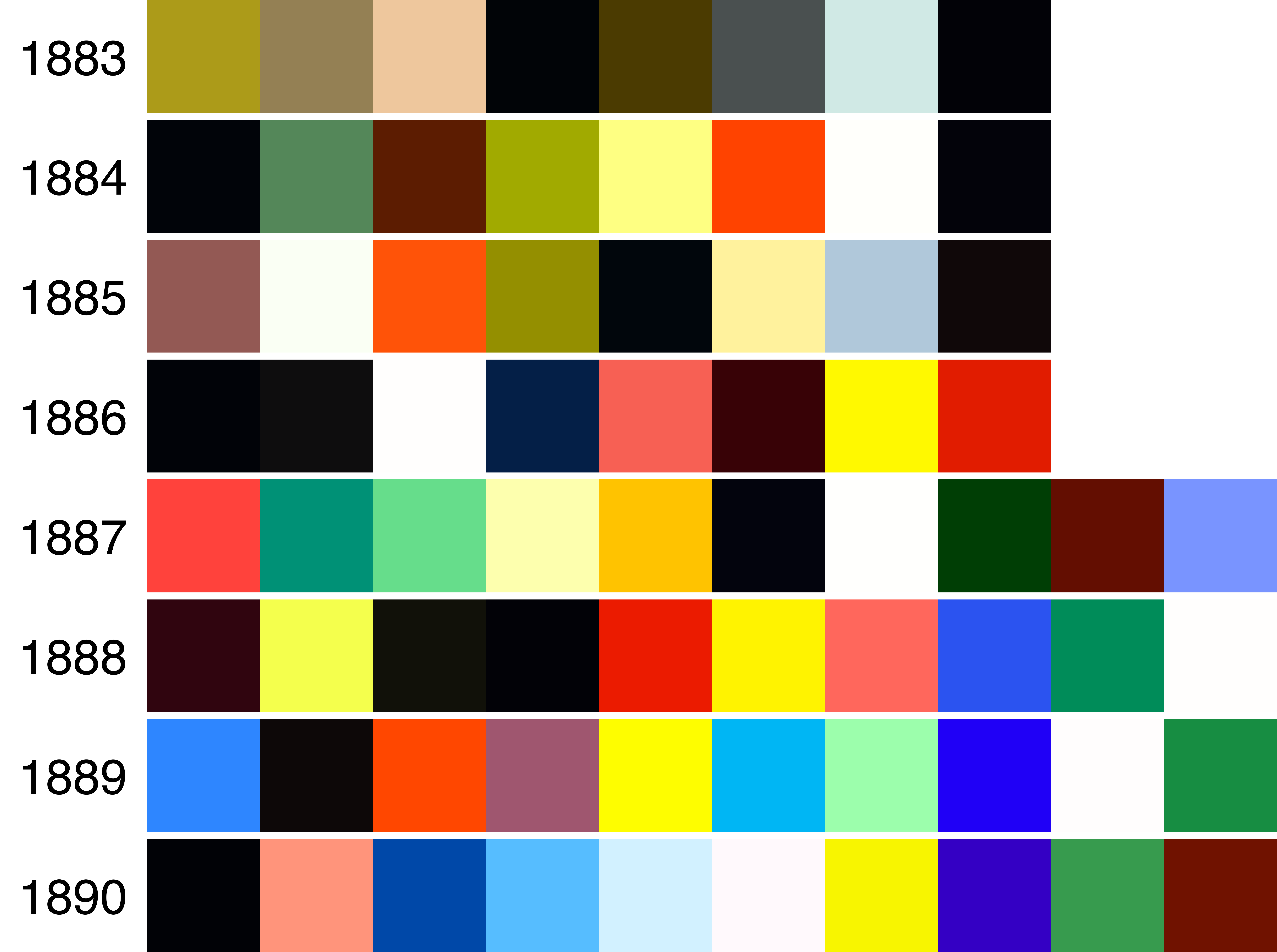}%
		\label{fig:VanGoghResults}
	}
	\caption{Examples of palette summarization.}
\end{figure}

The first stage of our algorithm can also be seen as yet another method for extracting a small palette from an arbitrary image, not necessarily of paintings.  Tan et al.~\shortcite{Tan:2016:DIL} and Chang et al.~\shortcite{Chang:2015:PPR} both present palette-extraction methods, as does Adobe's Color CC app~\shortcite{kuler}.  We compare these results in Fig.~\ref{fig:PhotoPaletteResults}, where it is clear to see that Chang et al.~\shortcite{Chang:2015:PPR} and Kuler attempt to find ``salient'' or meaningful colors in some sense, whereas Tan et al.~\shortcite{Tan:2016:DIL} and our work focus on colors that reconstruct the images.  We achieve similar results to Tan et al.~\shortcite{Tan:2016:DIL}, but as we showed earlier our reconstructions are much lower error for the same number of colors, as we have more success with paint-like color mixtures such as green and cyan.

We can also use our palette extraction method to analyze collections of images, by amending our method to jointly reconstruct the pixels of multiple images.  We use this approach to extract aggregate palettes from paintings of Van Gogh organized by year (Fig.~\ref{fig:VanGoghResults}).  Two results are clear from this analysis. First, the range of colors Van Gogh painted with expanded over the 1880's, as we expanded from eight pigments to ten pigments to achieve good reconstruction errors. Second, the vibrancy increased dramatically as well. 

\subsection{Edge Detection and Enhancement}

\begin{figure}
	\centering
	\includegraphics[width=0.92\columnwidth]{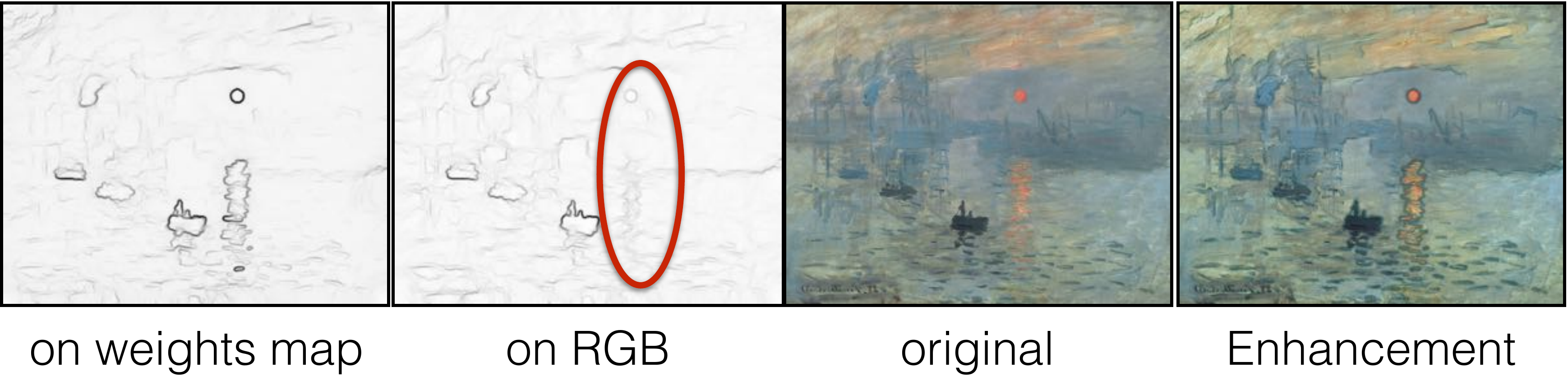}
	\caption{Paint-aware edge detection and enhancement.}
	\label{fig:boundaryEnhancement}
\end{figure}

Our weight maps can improve edge-based image analysis (Fig.~\ref{fig:boundaryEnhancement}).  We apply an existing edge detection method~\cite{isola2014crisp} to each weight map separately and merge the per-pigment response as the per-pixel max.  Paint edge images can be used to adapt standard image processing routines to be paint-aware.  For example, we do edge enhancement by thickening pigments near boundaries according to the edge response, which can visually emphasize painted objects in a different way than RGB edge enhancement.

%
%
%
%
%
%


\section{Conclusion}
\label{sec:conclusion}

We demonstrate a method that can recover plausible physical pigments from only an RGB image of a painting, and then recover the mixing proportion of those pigments at each pixel.  We are able to accurately reconstruct the RGB values of the image, and even closely match multispectral reflectance per-pixel as well, though the underlying pigment coefficients may differ.  We use this decomposition to enable a number of image editing operations that occur in ``pigment space,'' which creates results in a style more consistent with natural media imagery rather than digital RGB edits.

\paragraph{Limitations}
First, our approach requires users to choose a target number of primary pigments.
While this is the only user interaction in our entire pipeline, it is still a decision that the user must make. As shown in Fig.~\ref{fig:palette_size}, a large number of primary pigments will have lower RGB image reconstruction error at the cost of more tedious edits and additional processing time.
Second, if a ground truth primary pigment can be mixed from other primary pigments, our technique will not find it. However, this is an ambiguous situation, since it is not needed for perfect reconstruction. Still, it may be important for applications like pigment identification. 
Third, we use the Okumura~\shortcite{okumura2005developing} dataset as a prior
to help find initial \revtwo{$a,s$} values for our optimization, though our solutions are not limited to them.
We do not have other pigment datasets to verify whether this prior causes us to overfit our recovered pigment parameters. For example, this prior information may be more helpful for finding acrylic or oil pigments than watercolor.
Fourth, we assume constant pigment thickness. This is a simplifying assumption that speeds our optimization since the solution space is already under-constrained.
\rev{However, if we allowed for varying thickness, darker and lighter tones of a pigment could be obtained without a black or white pigment. We could reduce our palette size by computing
the 2D convex hull of the chromaticity of the pixels, ignoring brightness (Section~\ref{sec:method:estimating_primary_pigments}).
This is similar to the approach used by Aharoni-Mack et al.~\cite{aharoni2017pigment}.
We would also be better able to handle media like watercolor with varying translucency.
Finally, our approach does not recover ground truth, just plausible results enabling paint-like editing.}

\paragraph{Future work}
In the future, we would like to extend our result to estimate pigment layers instead of just mixtures. We plan to use our decomposition to help extract brush stroke-level structure from images of paintings, to enable manipulation of the brush strokes in painting images.
We predict that interpreting complex image structures with more appropriate models
will have applications in many applications of computer graphics.

\section{Acknowledgment}
\revthree{We thank the anonymous reviewers for their inspirational comments and suggestions. We are also grateful to Ya\u{g}{\i}z Aksoy for providing comparison results, and to the artists whose paintings we analyzed.
This work was supported in part by the United States National Science Foundation (IIS-1451198, IIS-1453018), a Google research award, and gifts from Adobe Systems, Inc.}



\bibliographystyle{IEEEtran}
\bibliography{bib/timemap.bib,bib/singlelayer.bib,bib/pigmento.bib}

\end{document}